\documentclass[preprint,showpacs,eqsecnum,aps,amsmath,amssymb,floatfix]{revtex4}
\usepackage{graphics}
\usepackage{epsfig}

\begin{document}

\title{Semiclassical unified description of wobbling motion in even-even and even-odd nuclei}

\author{A. A. Raduta$^{a), b)}$,  R. Poenaru $^{a)}$ and L. Gr. Ixaru $^{a), b)}$ }

\address{$^{a)}$ Department of Theoretical Physics, Institute of Physics and
  Nuclear Engineering, Bucharest, POBox MG6, Romania}

\address{$^{b)}$Academy of Romanian Scientists, 54 Splaiul Independentei, Bucharest 050094, Romania}

\begin{abstract}
A unitary description for wobbling motion in even-even and even-odd nuclei is presented. In both cases compact formulas for wobbling frequencies are derived. The accuracy of the harmonic approximation is studied for the yrast as well as for the excited bands in the even-even case.
Important results for the structure of the wave function and its behavior inside the two wells of the potential energy function corresponding to the Bargmann representation are pointed out.
Applications to $^{158}$Er and $^{163}$Lu reveal a very good agreement with available data.
Indeed, the yrast energy levels in the even-even case and the first four triaxial super-deformed bands,
TSD1,TSD2,TSD3 and TSD4, are realistically described. Also, the results  agree with the data for the E2 and M1 intra- as well as inter-band transitions. Perspectives for the formalism development
and an extensive application to several nuclei from various regions of the nuclides chart are presented.
\end{abstract}

\maketitle

\renewcommand{\theequation}{1.\arabic{equation}}
\setcounter{equation}{0}

\section{Introduction}
\renewcommand{\theequation}{1.\arabic{equation}}
\setcounter{equation}{0}
Many collective properties of the low lying states are related to the quadrupole collective coordinates. The simplest phenomenological scheme of describing them is the liquid drop model (LD) proposed by Bohr and Mottelson \cite{BohrMot}. Within the intrinsic frame of reference the liquid drop coordinates $\beta,\gamma, \Omega$ are described by a differential coupled equation from which one derives an uncoupled equation for the dynamical variable $\beta$ \cite{Ghe78,Rad78}. However, the rotational degrees of freedom, i.e. the Euler angles describing the position of the intrinsic frame with respect to the laboratory frame, and the variable $\gamma$, i.e. the deviation from the axial symmetry, are coupled together \cite{Rad78}. Under certain approximations 
\cite{Caprio} the equation describing the dynamic deformation $\gamma$ is separated from the one associated to the rotational degrees of freedom. Recently many papers were devoted to the study of the resulting equation for the gamma variable \cite{Iachello,Bona,Bona2,Bona3,Ghe07,Rad09} as well as the associated symmetries. Naturally since most of nuclei are axially symmetric these type of nuclei were intensively studied by both experimentalists and theoreticians.
However, the gamma degree of freedom is very important in determining many nuclear properties. This
justifies the attention payed to the $\gamma$ variable even in the early stage of nuclear structure
\cite{WilJean,Davy,TerVen}. An extensive study of the triaxial rotor and its coupling with the correlated individual degrees of freedom was achieved in 
Refs.\cite{Toki1,Toki2,Toki3,Toki4, Toki5,Tana}.The existence of a $\gamma$ deformed minimum in the potential energy surface leads to specific spectroscopic properties. One of the most exciting features of triaxial nuclei is their possible wobbling motion, which implies a precession of the total angular momentum combined with and oscillation of its projection on the quantization axis around a steady position. The first suggestion for a wobbling motion in nuclei was made by Bohr and Mottelson for high spin states in which the total angular momentum almost aligns to the principal axis with the largest moment of inertia, within the rotor model \cite{BMott}. A fully microscopic description of the wobbling phenomenon was achieved by Marshalek in Ref.\cite{Marsh}.
Since then a large volume of experimental as well as of theoretical results has been accumulated 
\cite{Odeg,Jens,Ikuko,Scho,Amro,Gorg,Ham,Matsu,Ham1,Jens1,Hage,Tana3,Oi,Bring,Hart,Cast,Alme}.
Experimentally, the wobbling states excited on the triaxial strongly deformed (TSD) bands are known
not only in $^{163,165,167}$Lu but also in $^{161}$Lu and $^{167}$Ta \cite{Bring,Hart}.
  
The main purpose of the present paper is to study the coupling of an individual nucleon to an even-even-core and apply the results to the description of the bands determined by the wobbling motion of the system. Aiming at a unified description of the wobbling motion in even-even and even-odd systems, we consider also the case of even-even nuclei. For this case the odd-spin sequence, belonging to the $\gamma$ band, is also considered. Indeed, according to Ref.\cite{MikIans} such a set of spins cannot be distinguished from the wobbling states through the $D_2$ symmetry and therefore it is natural to treat them on equal footing.
Although the even-even case was earlier treated in Ref. \cite{RaBuRa}, here we review the main ideas launched there and complete them with new theoretical results and numerical analysis.

The above sketched project will be accomplished in the following sections according to the following plan. In Section II the even-even system is treated. Quantitative comparison obtained in the quantum mechanical Hilbert space associated to the triaxial rotor, in the classical phase space and within the space spanned by the solutions of the Bargmann equation is presented. Numerical results for energies of the yrast states as well as of their electric and magnetic properties reflected in the E2, M1 transitions as well as in the electric quadrupole and magnetic dipole moments are presented for $^{158}$ Er. In Section III we treat an odd system consisting in a particle moving in a deformed mean-field and coupled to a triaxial rigid rotor. The system is dequantized via a variational principle and compact formulas for energies and reduced transition probabilities are derived. The numerical application refers to the isotope of 
$^{163}$Lu and results are compared with the available data. Final conclusions are drawn in Section IV. 

\section{New result for wobbling motion in even-even nuclei}
\renewcommand{\theequation}{2.\arabic{equation}}
\setcounter{equation}{0}
We suppose that some properties of triaxial nuclei can be quantitatively described by a triaxial rigid rotor.
Therefore, we  consider a  triaxial rigid rotor with the moments of inertia ${\cal I}_k$, k=1,2,3, corresponding to the axes  of the  intrinsic  frame, described by the Hamiltonian:
\begin{equation}
\hat{H}_{R}=\frac{\hat{R}^{2}_{1}}{2{\cal I}_{1}}+\frac{\hat{R}^{2}_{2}}{2{\cal I}_{2}}+\frac{\hat{R}^{2}_{3}}{2{\cal I}_{3}}.
\end{equation}
The  angular momentum components  are denoted by $\hat{R}_k$. 
This quantum mechanical object has been extensively studied in various contexts \cite{Casimir}, including that of nuclear physics \cite{Davy}. Indeed, in 
Ref.\cite{Davy}, the authors noticed that there are some nuclei whose low lying excitations might be described by the eigenvalues of a 
rotor Hamiltonian with suitable choice for the moments of inertia. Since then, many extensions of the rotor picture have been considered. We just mention few of them: particle-rotor model \cite{TerVen}, two rotors model \cite{Iudi} used for describing the scissors modes, the cranked triaxial rotor \cite{Ghe}.
The extensions provide a simple description of the data but also lead to new findings like scissors mode \cite{Iudi}, finite magnetic bands, chiral symmetry \cite{Fra}.

In principle it is easy to find the eigenvalues of $H_R$ by using a diagonalization procedure within a basis exhibiting the $D_2$ symmetry. However, when we restrict the considerations to the yrast band it is by far more convenient  to use a closed expression for the excitation energies.

We suppose that a certain class of properties of the Hamiltonian $H_R$ can be obtained by solving the time dependent equations provided by the variational principle:

\begin{equation}
\delta \int_{0}^{t} \langle \psi (z)|H-i\frac{\partial}{\partial t^{\prime}}|\psi (z)\rangle dt^{\prime}=0.
\label{varec}
\end{equation}
If the trial function $|\psi (z)\rangle$ spans the whole Hilbert space of the wave functions describing the system, solving the equations provided by the variational principle is equivalent to solving the Schr\"{o}dinger equation associated to $H_R$. Here we restrict the Hilbert space to the subspace spanned by the the variational state:
\begin{equation}
|\psi(z)\rangle ={\cal N}e^{z\hat{R}_-}|IMI\rangle ,
\label{trial}
\end{equation}
where $z$ is a complex number depending on time and $|IMK\rangle $ denotes the eigenstates of the angular momentum operators $\hat{R}^2$, $R_z$ and $\hat{R}_3$ with $R_z$ denoting the angular momentum projection on the OZ axis of the laboratory frame. ${\cal N}$ is a factor which assures that the function $|\psi\rangle$ is normalized to unity: 
\begin{equation}
{\cal N}=(1+|z|^{2})^{-I}.
\end{equation}
$\hat{R}_-$ denotes the lowering operator which for the intrinsic components is :
\begin{equation}
\hat{R}_-=\hat{R}_1+i\hat{R}_2.
\end{equation}
The function $(\ref{trial})$ is a coherent state for the group $SU(2)$ \cite{Kura}, generated by the angular momentum components, and is suitable for the description of the classical features of the rotational degrees of freedom. Due to the supercompletness property, the variational state comprises all basis vectors spanning the Hilbert space. Actually this is the feature which assures
a good approach to the eigenfunctions of $H$. As a matter of fact this will be concretely checked out within the present section.
\subsection{Canonical conjugate coordinates}

The averages of $H_R$ and the time derivative operator with the function (2.3), have the expressions:
\begin{eqnarray}
\langle\hat{H}\rangle &=&\frac{I}{4}\left(\frac{1}{{\cal I}_{1}}+\frac{1}{{\cal I}_{2}}\right)+\frac{I^{2}}
{2{\cal I}_{3}}+\frac{I(2I-1)}{2(1+zz^{*})^{2}}\left[\frac{(z+z^{*})^{2}}{2{\cal I}_{1}}-\frac{(z-z^{*})^{2}}{2{\cal I}_{2}}-\frac{2zz^{*
}}{{\cal I}_{3}}\right], \nonumber\\
\langle\frac{\partial}{\partial{t}}\rangle &=&\frac{I(\stackrel{\bullet}{z}z^{*}-z
\stackrel{\bullet}{z}^{*})}{1+zz^{*}}.
\end{eqnarray}
Denoting the average of $H_R$ by ${\cal H}$, the time dependent variational equation yields:
\begin{equation}
\frac{\partial{\cal{H}}}{\partial{z}}=-\frac{2iI\stackrel{\bullet}{z}^{*}}{(1+zz^{*})^{2}},\;\;
\frac{\partial{\mathcal{H}}}{\partial{z^{*}}}=\frac{2iI\stackrel{\bullet}{z}}{(1+zz^{*})^{2}}.
\end{equation}

In terms of the  polar coordinate ( $z=\rho e^{i\varphi}$ ), the equations of motion become:
\begin{equation}
\frac{\partial{\mathcal{H}}}{\partial{\rho}}=-\frac{4\rho I\stackrel{\bullet}{\varphi}}{(1+\rho^{2})^{2}},\;\;
\frac{\partial{\mathcal{H}}}{\partial{\varphi}}=\frac{4I\rho\stackrel{\bullet}{\rho}}{(1+\rho^{2})^{2}}.
\end{equation}
The pair of conjugate variables which brings the classical equations of motion in the canonical Hamilton form is $(r,\varphi)$ with $r$ having the expression:
\begin{equation}
r=\frac{2I}{1+\rho^2},\;\;0\le r\le 2I.
\end{equation}
Indeed, in the new variables the equations of motion are:
\begin{equation}
\frac{\partial{\cal{H}}}{\partial{r}}=\stackrel{\bullet}{\varphi},\;\;
\frac{\partial{\cal{H}}}{\partial{\varphi}}=-\stackrel{\bullet}{r}.
\label{ecmot}
\end{equation}
Accordingly, $\varphi$ and $r$ play the role of generalized coordinate and momentum respectively.
The classical energy function acquires the expression:
\begin{equation}
{\cal H}(r,\varphi)=\frac{I}{4}\left(\frac{1}{{\cal I}_{1}}+\frac{1}{{\cal I}_{2}}\right)+\frac{I^{2}}{2{\cal I}_{3}}+\frac{(2I-1)r(2I-r)}{4I}\left[\frac{\cos^{2}{\varphi}}{{\cal I}_{1}}+\frac{\sin^{2}{\varphi}}{{\cal I}_{2}}-\frac{1}{{\cal I}_{3}}\right].
\end{equation}

Averaging the angular momentum components with the function $|\psi(z)\rangle$ one obtains:
\begin{equation}
\langle I_1\rangle=\frac{2I\rho}{1+\rho^2}\cos\varphi,\;\;\langle I_2\rangle =\frac{2I\rho}{1+\rho^2}\sin\varphi,\;\;\langle I_3\rangle =I\frac{1-\rho^2}{1+\rho^2}.
\end{equation}
Another pair of canonically conjugate coordinates is:
\begin{equation}
\xi=I\frac{1-\rho^2}{1+\rho^2}=\langle I_3\rangle \; \rm{and}\;\phi=-\varphi,
\end{equation} 
Indeed, their equations of motion are:
\begin{equation}
\frac{\partial{\cal H}}{\partial \xi} =-\stackrel{\bullet}{\phi},\;\;
\frac{\partial{\cal H}}{\partial \phi} =\stackrel{\bullet}{\xi}.
\end{equation}
Taking the Poisson bracket defined in terms of the new conjugate coordinates one finds:
\begin{equation}
\{\langle I_1\rangle ,\langle I_2\rangle \}=\langle I_3\rangle, \;\;
\{\langle I_2\rangle ,\langle I_3\rangle \}=\langle I_1\rangle,\;\;
\{\langle I_3\rangle ,\langle I_1\rangle \}=\langle I_2\rangle
\end{equation}
Therefore  the angular momentum components form a classical algebra, $SU(2)_{cl}$, with the inner product $\{ , \}$. The correspondence
\begin{equation}
\{\langle I_k \rangle ,\{ , \}i\}\longrightarrow \{I_k, [ , ]\},
\end{equation}
is an isomorphism of $SU(2)$ algebras, which accomplishes the quantization of the classical  angular momentum.

The dequantization method is very useful when in the classical picture some confident approximations can be performed and then the classical trajectories are quantized. In most cases the transition from the Hilbert space associated to the initial quantal Hamiltonian to the classical phase space  and from the classical phase space to the new quantum mechanical Hilbert space, through a quantization procedure, are achieved with some inherent approximations. In this context it is necessary to compare the quantum result with those obtained in the classical phase space as well as with those obtained in the re-quantized picture. A general answer for a quantitative comparison between the results obtained within the three spaces i.e. the Hilbert space, the phase space and the space of re-quantized trajectories  is not yet available. However, it is known that solving the classical equations is equivalent to solving the initial time dependent Schr\"{o}dinger equation if the trial function spans the whole Hilbert space. Moreover, even the classical equations cannot be solved without adopting some specific approximations.
The aim of this Section is to show that for the triaxial rotor the three classes of results agree with each other impressively well. 

The classical trajectories are quantized by changing the real conjugate variables to a pair of complex canonical conjugate variables and then identifying these with a pair of a creation and a annihilation boson operator, respectively. According to the Darboux theorem \cite{Cartan}, the pair of canonical complex coordinate is not unique. Moreover, the quoted theorem provides a recipe for finding new pairs of conjugate variables.  For the case treated here, we suggested four pairs of canonical complex coordinates \cite{RaBuRa}. For the present goals we mention only two of them, namely those which by quantization lead to the well known Holstein-Primakoff and Dyson boson representation (alternatively called as {\it boson expansion}) of the quantum angular momentum algebra.

To begin with, let us consider the average of the angular momentum components, expressed in terms of the variables ($\varphi, r$):
\begin{eqnarray}
J^{cl}_+\equiv \langle \hat{I}_{+}\rangle &=&\sqrt{r(2I-r)}\cdot e^{i\varphi},\nonumber\\
J^{cl}_-\equiv \langle \hat{I}_{-}\rangle &=&\sqrt{r(2I-r)}\cdot e^{-i\varphi},\nonumber\\
J^{cl}_3\equiv \langle \hat{I}_{3}\rangle &=&I-(2I-r)=r-I.
\end{eqnarray}

\subsection{Complex coordinates and their quantization} 
The Poisson bracket associated to  any two complex functions defined on the classical phase space is defined by:
\begin{equation}
\{f,g\}=\frac{\partial f}{\partial \varphi}\frac{\partial g}{\partial r} -\frac{\partial f}{\partial r}\frac{\partial g}{\partial \varphi}.
\end{equation}
With this definition the equations of motion of the classical coordinates may be written as::
\begin{equation}
\{r,{\cal H} \}=\dot{r},\;\; \{ \varphi, {\cal H} \}=\dot{\varphi},\; \{ \varphi, r\}=1.
\end{equation}
The classical angular momentum components satisfy the equations:
\begin{equation}
\{J^{cl}_+,J^{cl}_-\}=-2iJ^{cl}_3,\; \{{J^{cl}}_{\pm},J^{cl}_3\}=\pm i{J^{cl}}_{\pm}.
\end{equation}
The functions $J^{cl}_{\pm}, J^{cl}_3$ with the inner product defined by the Poisson brackets,  generate a classical algebra which will be denoted by $SU_{cl}(2)$.
\subsection{Holstein-Primakoff boson expansion}
Let us consider the complex coordinate
\begin{equation}
{\cal C}=\sqrt{2I-r}\cdot e^{-i\varphi},
\end{equation}
and denote by ${\cal C}^*$ the corresponding complex conjugate variable.
They obey the equations:
\begin{equation}
\{{\cal C},{\cal C}^{*}\}=i,\;\;\{{\cal C},{\cal H}\}=\stackrel{\bullet}{{\cal C}},\;\;\{{\cal C}^*,{\cal H}\}=\stackrel{\bullet}{{\cal C}}^*.
\end{equation}
These equations suggest that the complex coordinates are of canonical type.
To quantize the classical phase space means to achieve a homeomorphism between the algebra of the  ${\cal C}, {\cal C}^*$ complex functions, with the multiplication operation $\{,\}$ and the algebra of the boson operators $a,a^{\dagger}$, with the commutator as inner multiplier:
\begin{equation}
\left({\cal C},{\cal C}^*,\{,\}\right)\longrightarrow \left(a, a^{\dagger},-i[,]\right).
\end{equation}

The quantization of an arbitrary function $f({\cal C},{\cal C}^*)$ is performed by replacing ${\cal C}$ and ${\cal C}^*$ by the operators $a$ and $a^{\dagger}$, respectively. Concerning the terms containing mixed products of ${\cal C}$ and ${\cal C}^*$, these must be symmetrized first and then the complex coordinates be replaced by the boson operators.
The simplest example is the angular momentum components which after quantization become:
\begin{eqnarray}
\hat{J}_{+}&=&\sqrt{2I}\,a^{\dagger}\left(1-\frac{a^{\dagger}a}{2I}\right)^{\frac{1}{2}},
\nonumber\\
\hat{J}_{-}&=&\sqrt{2I}\,\left(1-\frac{a^{\dagger}a}{2I}\right)^{\frac{1}{2}}a,\nonumber\\
\hat{J}_{3}&=&I-a^{\dagger}a.
\label{HoPr}
\end{eqnarray}
One can check that these boson operators obey the specific commutation relations of the intrinsic angular momentum components and, consequently, generate an $SU(2)$ algebra which hereafter will be denoted by $SU_b(2)$.
The product of the two successive homeomorphisms:
\begin{equation}
SU(2)\to SU_{cl}(2)\to SU_{b}(2)
\end{equation}
is a homeomorphism $SU(2)\to SU_b(2)$ which, in fact, is the boson representation of the angular momentum algebra. The equations 
 (\ref{HoPr}) are known under the name of Holstein and Primakoff {\it boson expansion} (HP) for the angular momentum components \cite{HolPr}.
\subsection{Dyson boson expansion}
The pair of canonical complex variables which generates the Dyson's {\it boson expansion} (D) for the angular momentum is.
\begin{eqnarray}
{\cal C}_1&=&\sqrt{2I}\sqrt{\frac{2I-r}{r}} e^{-i\varphi},\nonumber\\
{\cal B}^*_1&=&\frac{1}{\sqrt{2I}}\sqrt{r(2I-r)} e^{i\varphi}.
\end{eqnarray}
Indeed, their Poisson bracket is:
\begin{equation}
\{{\cal B}^*_1,{\cal C}_1\}=i.
\end{equation}
In the next step, the complex coordinates are quantized
\begin{equation}
\left({\cal C}_1,{\cal B}^*_1,\{,\}\right)\longrightarrow \left(b, b^{\dagger},-i[,]\right),
\end{equation}
and thus the Dyson's boson representation (D) of angular momentum \cite{Dys} is obtained:
\begin{eqnarray}
\hat{J}^{D}_{+}&=&\sqrt{2I}b^{\dagger}\nonumber\\
\hat{J}^{D}_{-}&=&\sqrt{2I}\left(1-\frac{b^{\dagger}b}{2I}\right)b,\nonumber\\
\hat{J}^{D}_{3}&=&I-b^{\dagger}b. 
\end{eqnarray}
Note that while the HP expansion preserves the hermiticity property, the D expansion does not have such a virtue.

\subsection{Harmonic approximation for the energy function}
 Suppose we solved the classical equations of motion (\ref{ecmot}) and the classical trajectories given by $\varphi =\varphi(t),\; r=r(t)$ are found.
Due to Eq.(\ref{ecmot}), one finds that the time derivative of ${\cal H}$ is vanishing. This means that the system energy is a constant of motion and, therefore, the trajectory lies on the surface ${\cal H} =const.$. Another restriction for trajectory consists in the fact that the classical angular momentum squared is equal to $I(I+1)$. This restriction is automatically fulfilled by the classical angular momentum. The intersection of the two surfaces, defined by the two constants of motion, 
determines the manifold to which the system trajectory belongs.

Studying the sign of the Hessian associated to ${\cal H}$, one obtains the points where ${\cal H}$ acquires extremal values. Here we consider only the case 
${\cal I}_{1}>{\cal I}_{3}>{\cal I}_{2}$, \, when $(0,I)$ is a minimum point for energy, while 
$(\frac{\pi}{2}, I)$ a maximum.

The second order expansion for ${\mathcal{H}}(r,\varphi)$ around the minimum point, yields:
\begin{equation}
\tilde{\mathcal{H}}(r,\varphi)=\frac{I}{4}\Bigg(\frac{1}{{\cal I}_{2}}+\frac{1}{{\cal I}_{3}}\Bigg)+\frac{I^{2}}{2{\cal I}_{1}}+\frac{2I-1}{4I}\Bigg(\frac{1}{{\cal I}_{3}}-\frac{1}{{\cal I}_{1}}\Bigg)r^{\prime 2}+\frac{(2I-1)I}{4}\Bigg(\frac{1}{{\cal I}_{2}}-
\frac{1}{{\cal I}_{1}}\Bigg)\varphi^{\prime 2}.
\end{equation}
This equation describes an oscillator with the frequency:
\begin{equation}
\omega_I=\left(I-\frac{1}{2}\right)\sqrt{\Bigg(\frac{1}{{\cal I}_{3}}-\frac{1}{{\cal I}_{1}}\Bigg)\Bigg(\frac{1}{{\cal I}_{2}}-\frac{1}
{{\cal I}_{1}}\Bigg)}.
\end{equation}
This frequency is associated to the precession motion of the angular momentum around the OX axis.
In our description, the yrast band energies are, therefore, given by:  
\begin{equation}
E_I=\frac{I}{4}\Bigg(\frac{1}{{\cal I}_{2}}+\frac{1}{{\cal I}_{3}}\Bigg)+\frac{I^{2}}{2{\cal I}_{1}}
+\frac{\omega_I}{2}.
\label{yra}
\end{equation}

\subsection{The Bargmann representation}
The classical energy function can be quantized by one of the two procedures mentioned above. In particular, by expressing ${\cal H}$ in terms of the complex conjugate variable $C_1$ and $B^*_1$
and then replacing these by the bosons $b$ and $b^+$, one obtains the Dyson boson representation of $H_R$ denoted hereafter by $H_D$. Although  this boson operator is not Hermitian it has real eigenvalues \cite{Ogu}. We searched for the eigenvalues of $H_D^{\dagger}$ by using the Bargmann representation of the boson operators \cite{Bar,Janc,Jan}:
\begin{equation}
b^{\dagger}\to x,\;\;b\to\frac{d}{dx}
\label{Bquant}
\end{equation}
In this way the eigenvalue equation of $H_D^{\dagger}$ is transformed into a differential equation:

\begin{equation}
\left[\left(-\frac{k}{4I}x^4+x^2-kI\right)\frac{d^2}{dx^2}+(2I-1)\left(\frac{k}{2I}x^3-x\right)\frac{d}{dx}-k\left(I-\frac{1}{2}\right)x^2\right]G=E^{\prime}G.
\label{ecdif}
\end{equation}
where
\begin{equation}
k=\frac{\frac{1}{{\cal I}_1}-\frac{1}{{\cal I}_2}}{\frac{1}{{\cal I}_1}+\frac{1}{{\cal I}_2}-
\frac{2}{{\cal I}_3}}.
\end{equation}
It can be easily proved that this equation can be brought to  the algebraic form of the 
Lam\'{e} equation \cite{Bate,Byer}. 
Performing now the change of function and variable:
\begin{eqnarray}
G&=&\left(\frac{k}{4I}x^4-x^2+kI\right)^{I/2}F,\nonumber\\
t&=&\int_{\sqrt{2I}}^{x}\frac{dy}{\sqrt{\frac{k}{4I}y^4-y^2+kI}},
\label{tfuncx}
\end{eqnarray}
Eq.(\ref{ecdif}) is transformed into a second order differential Schr\"{o}dinger equation:
\begin{equation}
-\frac{d^2F}{dt^2}+V(t)F=E^{\prime}F,
\label{Schr}
\end{equation}
with
\begin{equation}
V(t)=\frac{I(I+1)}{4}\frac{\left(\frac{k}{I}x^3-2x\right)^2}{\frac{k}{4I}x^4-x^2+kI}-k(I+1)x^2+I.
\label{potent}
\end{equation}

The considered ordering for the moments of inertia is such that $k>1$. Under this circumstance the potential $V(t)$ has two minima for $x=\pm\sqrt{2I}$, and a maximum for x=0. 

The minimum value for the potential energy is:
\begin{equation}
V_{min}=-kI(I+1)-I^2.
\end{equation}
Note that the potential is symmetric in the variable x. Due to this feature the potential behavior  around the two minima are identical.
To illustrate the potential behavior around its minima we make the option for the minimum $x= \sqrt{2I}$. To this value of $x$ it corresponds, $t=0$. Expanding $V(t)$ around $t=0$ and truncating the expansion at second order we obtain:
\begin{equation}
V(t)=-kI(I+1)-I^2+2k(k+1)I(I+1)t^2.
\end{equation}
Inserting this expansion in Eq.(\ref{Schr}), one arrives at a Schr\"{o}dinger equation for an oscillator. The eigenvalues are
\begin{equation}
E_n^{\prime}=-kI(I+1)-I^2+\left[2k(k+1)I(I+1)\right]^{1/2}(2n+1).
\end{equation}  
The quantized Hamiltonian associated to ${\cal H}$, i.e. $H^{\dagger}_{D}$, has an eigenvalue which is obtained  from the above expression. The final result is:
\begin{equation}
E_{n,I}=\frac{I(I+1)}{2{\cal I}_1}+\hbar\omega_I(n+\frac{1}{2}).
\label{wfor}
\end{equation}
where 
\begin{equation}
\omega_I = \left[\left(\frac{1}{{\cal I}_2}-\frac{1}{{\cal I}_1}\right)\left(\frac{1}{{\cal I}_3}-\frac{1}{{\cal I}_1}\right)I(I+1)\right]^{1/2},
\end{equation}
defines the wobbling frequency of the angular momentum.

The Bargmann representation of the angular momentum components is obtained by inserting the correspondence (\ref{Bquant})
into the Dyson boson expansion. The result is:
\begin{eqnarray}
I_+&=&\sqrt{2I}x,\nonumber\\
I_-&=&\sqrt{2I}(\frac{d}{dx}-\frac{x}{2I}\frac{d^2}{dx^2}),\nonumber\\
I_0&=&I-x\frac{d}{dx}.
\end{eqnarray}
From these expressions one may derive the angular momentum component $I_1$, which may be further averaged with the wave function provided by the Schrodinger equation for a given value of I. As a result one obtains a maximal value ($=I$), which in fact confirms the result we got at the classical level.

It is instructive to compare the K-amplitudes of the yrast states obtained through diagonalization,
$A^{diag}_{K}$,  and those corresponding to the coherent state (2.3) considered in the minimum point $(\varphi,r)=(0,I)$ for I=20. The latter function can be written in a different form:
\begin{equation}
|\Phi_{IM}\rangle=\left.|\Psi_{IM}\rangle\right|_{0,I}=\sum_{K}\frac{1}{2^I}\left(\begin{matrix}2I\cr I-K\end{matrix}\right)^{1/2}|IMK\rangle \equiv\sum_{K}A^{coh}_K|IMK\rangle.
\label{Phi}
\end{equation}
The two sets of amplitudes were plotted in Fig. 1,
from where we see that the two functions have a similar K dependence. The small difference is caused by the fact that diagonalization provides non-vanishing amplitudes only for $K=even$, while the coherent state  comprises all K-components. The former state is degenerate with the second yrast state which has only $K=odd$ components. Combining the two degenerate functions to a normalized function, the $K$-distribution of the new function is almost identical to that of 
$|\Phi_{IM}\rangle$.  
\begin{figure}[ht!]
\includegraphics[width=0.6\textwidth]{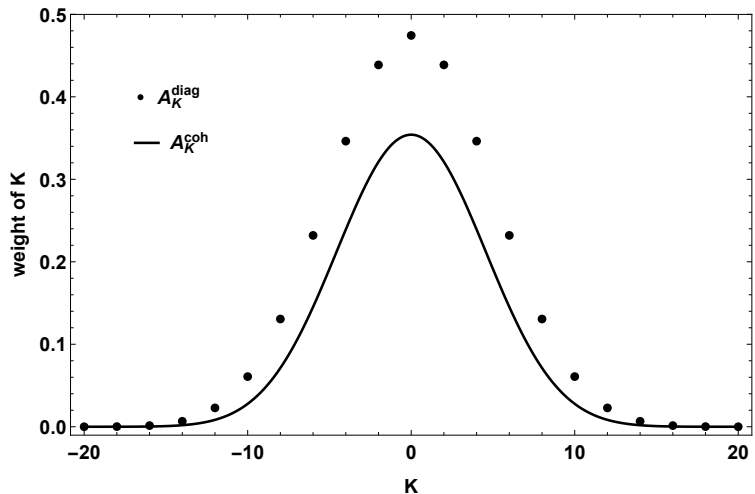}
\caption{The $K$-amplitudes supplied by the diagonalization procedure and the coherent state, respectively.}
\label{Fig.1}
\end{figure}

\subsection{Transition probabilities}
The transition operators for electric quadrupole and magnetic dipole transitions are:
\begin{eqnarray}
{\cal M}(E2;\mu)&=&\frac{3}{4\pi}Ze_{eff}R_0^2\left(D^2_{\mu 0}\beta\cos\gamma+(D^2_{\mu 2}+D^2_{\mu , -2})\beta\sin\gamma /\sqrt{2}\right),\nonumber\\
{\cal M}(M1;\mu)&=&\sqrt{\frac{3}{4\pi}}g_RD^1_{\mu\nu} R_{\nu}.
\label{troper}
\end{eqnarray}
Here Z and $R_0$ denote the nuclear charge and radius respectively, while $D^I_{MK}$ stands for the Wigner function describing the rotation matrix, $\mu_N$ is the nuclear magneton and $e_{eff}$-
the effective charge. $\beta$ is the nuclear deformation and $\gamma$ represents  the nuclear shape  deviation from the axial symmetry. They are not dynamic variables but real numbers fixed in the manner described in the next sub-section.

When the intra-band transition is concerned, the initial and final states of the yrast band are described by the function defined in Eq. (\ref{Phi}). Since the one  phonon operator of the yrast band is associated with the quantas in the parameter space, it commutes with the transition operator and moreover gives zero when acts on the final yrast state, unless this deviates from 
(\ref{Phi}) due to parameter fluctuations.
The normalized first order expansion of $\Psi$ around $\Phi$ is:
\begin{eqnarray}
|\Psi_{I+1,M}\rangle &=&N_{I+1}\frac{1}{2^I}\sum_{K=-I}^{K=I}\left[1+\frac{i}{\sqrt{2}}\left(\frac{\alpha_I K}{I}
+\frac{I-K}{\alpha_I}\right)\right]\left(\begin{matrix} 2I \cr I-K\end{matrix}\right)^{1/2}|IMK\rangle a_{I}^{\dagger}|0\rangle_I,
\;I\ne 0,\nonumber\\
\left(N_{I+1}\right)^{-2}&=&\frac{1}{2^{2I}}\sum_{K=-I}^{K=I}\left[1+\frac{1}{2}\left(\frac{\alpha_I K}{I}+\frac{I-K}{\alpha_I}\right)^2\right] \left(\begin{matrix} 2I \cr I-K\end{matrix}\right).
\end{eqnarray}
Here $a_{I}^{\dagger}$ denotes the creation operator for a wobbling quanta on the top of the yrast state of angular momentum I. The corresponding vacuum state is $|0\rangle_{I}$. The canonical transformation relating the conjugate coordinate and momentum with the creation and annihilation operators depend on the parameter $\alpha_I$ having the expression:
\begin{equation}
\alpha_I=\left(I^2\frac{\frac{1}{{\cal J}_2}-\frac{1}{{\cal J}_1}}{\frac{1}{{\cal J}_3}-\frac{1}{{\cal J}_1}}\right)^{1/4}.
\end{equation}
For what follows we introduce the following notation for a wobbling multi-phonon state:
\begin{equation}
|\Phi_{I+n_w,M};n_w\rangle = |\Phi_{IM}\rangle \frac{\left(a_{I}^{\dagger}\right)^{n_w}}{\sqrt{n_{w}!}}|0\rangle_I.
\end{equation}
The quadrupole transition amplitude is given by the reduced matrix element which, in the Rose's convention, is:
\begin{eqnarray}
&&\langle \Phi_I;n_w||{\cal M}(E2)||\Phi_{I'};n^{'}_{w}\rangle =\frac{3}{4\pi}ZR_0^2e_{eff}
\frac{1}{2^{I+I'}}\frac{\hat{I'}}{\hat{I}}\frac{i}{\sqrt{2}}\sum_{k,k'}
F_{n_{w}n^{'}_{w}}(K')\left(\begin{matrix} 2I \cr I-K\end{matrix}\right)^{1/2}
\left(\begin{matrix} 2I' \cr I'-K'\end{matrix}\right)^{1/2}\nonumber\\
&\times&\left[\beta\cos\gamma C^{I'\;2\;I}_{K\;0\;K}\delta_{K',K}+\frac{\beta\sin\gamma}{\sqrt{2}}
\left(C^{I'\;2\;I}_{K'\;2\;K}\delta_{K',K-2}+C^{I'\;2\;I}_{K'\;-2\;K}\delta_{K',K+2}\right)\right],
\nonumber\\
&&F_{n_{w}n^{'}_{w}}(K')=\delta_{n_{w},n^{'}_{w}}+\frac{i}{\sqrt{2}}\left(\frac{\alpha_{I'}K'}{I'}
+\frac{I'-K'}{\alpha_{I'}}\right)\delta{n^{'}_{w},n_{w-1}}(1-\delta_{I',0}).
\label{amptr}
\end{eqnarray}
The reduced transition probability is readily obtained:
\begin{equation}
B(E2;In_w\to I'n^{'}_w)=\left|\langle \Phi_I;n_w||{\cal M}(E2)||\Phi_{I'};n^{'}_{w}\rangle\right|^2.
\end{equation}
The transition amplitude (\ref{amptr})  can be also used to calculate the quadrupole moment of an yrast state of angular momentum I:
\begin{equation}
Q_{I}=\sqrt{\frac{16\pi}{5}}C^{I\;2\;I}_{I\;0\;I}\langle \Phi_{II}|{\cal M}(E2)|\Phi_{II}\rangle.
\end{equation}
The magnetic properties were studied with the dipole transition operator defined by 
Eq.(\ref{troper}). The result for the magnetic dipole moment for an yrast state $I$ is:
\begin{equation}
\mu_I\equiv\sqrt{\frac{4\pi}{3}}\langle\Phi_{II}|R_0|\Phi_{II}\rangle =\sqrt{\frac{4\pi}{3}}g_RC^{I\;1\;I}_{I\;0\;I}\sqrt{R(R+1)}\mu_N.
\end{equation}
where the standard notation, $\mu_N$, for the nuclear magneton has been used.

These expressions for the electric and magnetic transition operator matrix elements will be used in the next subsection to calculate the corresponding observables for the case of $^{158}$Er.

\subsection{Numerical analysis}
Here we address the issue of how do the results obtained through diagonalization, by solving the Schr\"{o}dinger equation and by the harmonic approximation leading to the wobbling motion of the angular momentum respectively, compare with each other. Since this analysis has a pure pedagogical character, we chose for the moments of inertia arbitrary values:

 $${\cal I}_1=125\hbar^2MEV^{-1},\; {\cal I}_2=31.4\hbar^2MEV^{-1},\; {\cal I}_3=42 \hbar^2MEV^{-1}$$. 
However, when we aim to describe the experimental energies, a fitting procedure for the moments of inertia will be adopted.
The variable x from the Bargmann representation of the rotor Hamiltonian is defined in the interval $(-\infty,+\infty)$, while the current variable $t$ entering the Schr\"{o}dinger equation is restricted in a finite interval which is close to $[-1.5,+1.5]$.
\begin{figure}[t!]
\includegraphics[width=0.5\textwidth]{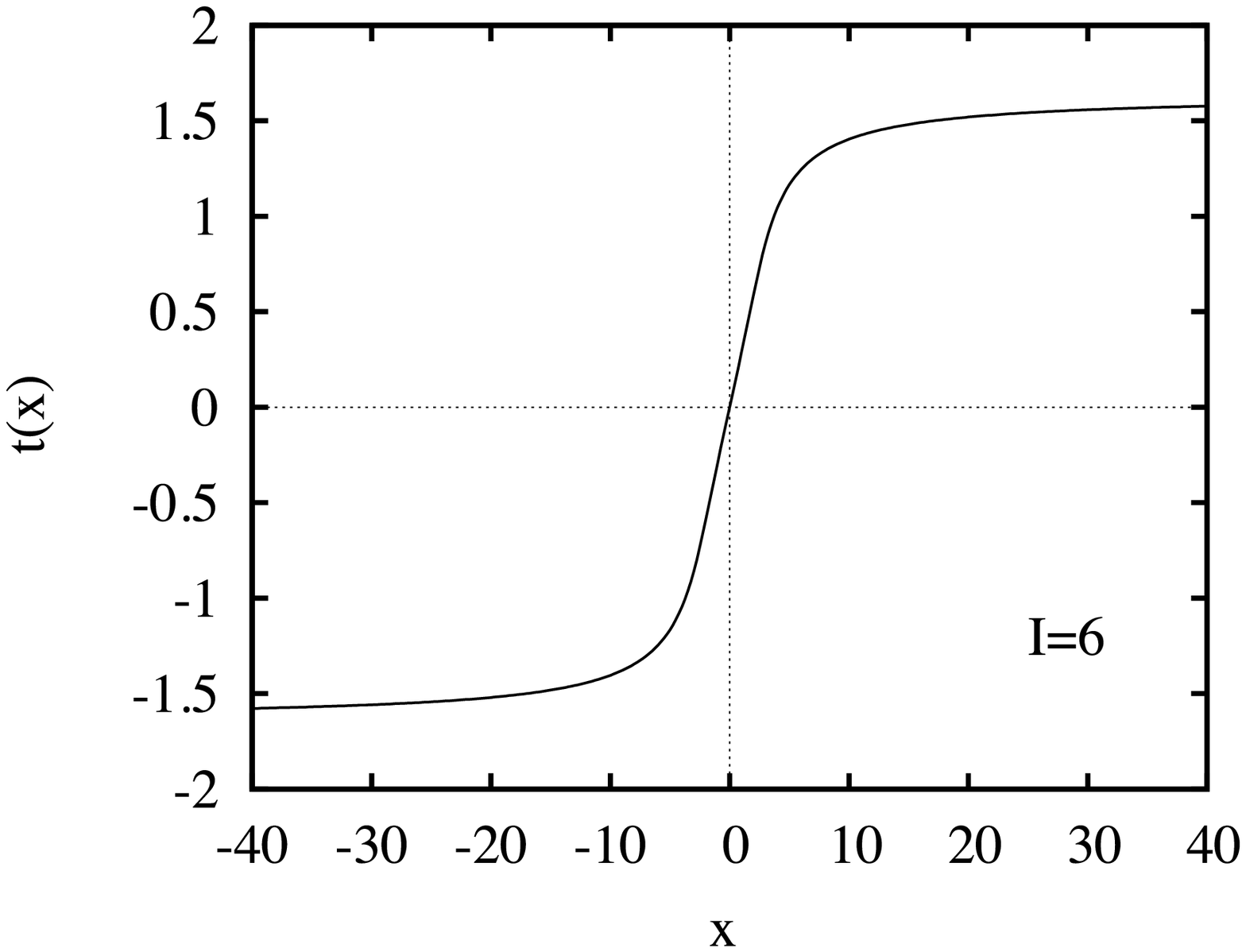}\includegraphics[width=0.5\textwidth]{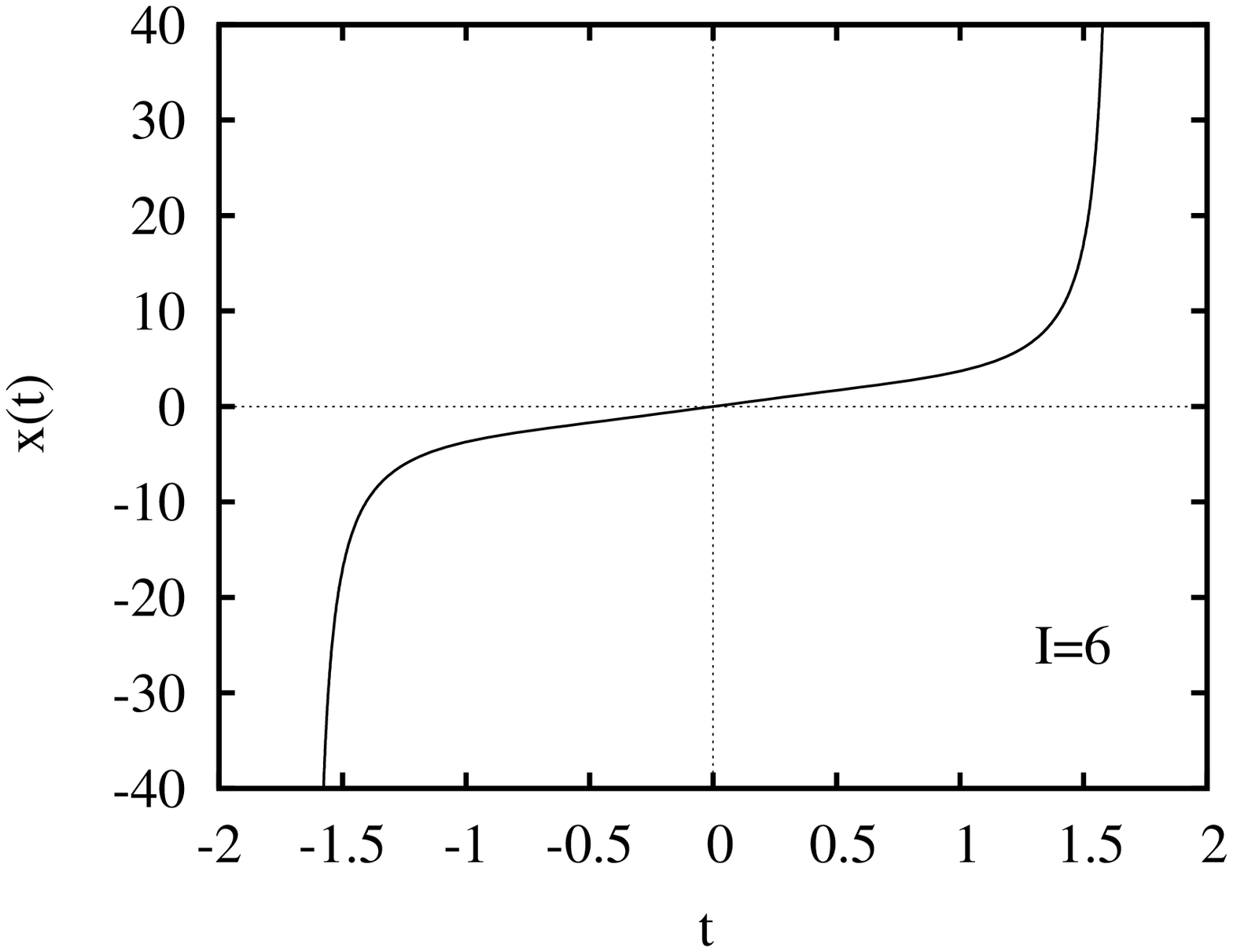}
\begin{minipage}{7.5cm}
\caption{The variable t defined by Eq. (\ref{tfuncx}) as a function of x defined by Eq. (2.33)}
\label{Fig. 2}
\end{minipage}\ \
\begin{minipage}{7.5cm}
\caption{The variable x, defined by Eq.(2.33),  as function of x given by 
Eq.(\ref{tfuncx}) }
\label{Fig.3}
\end{minipage}
\end{figure}
\begin{figure}[t!]
\includegraphics[width=0.5\textwidth]{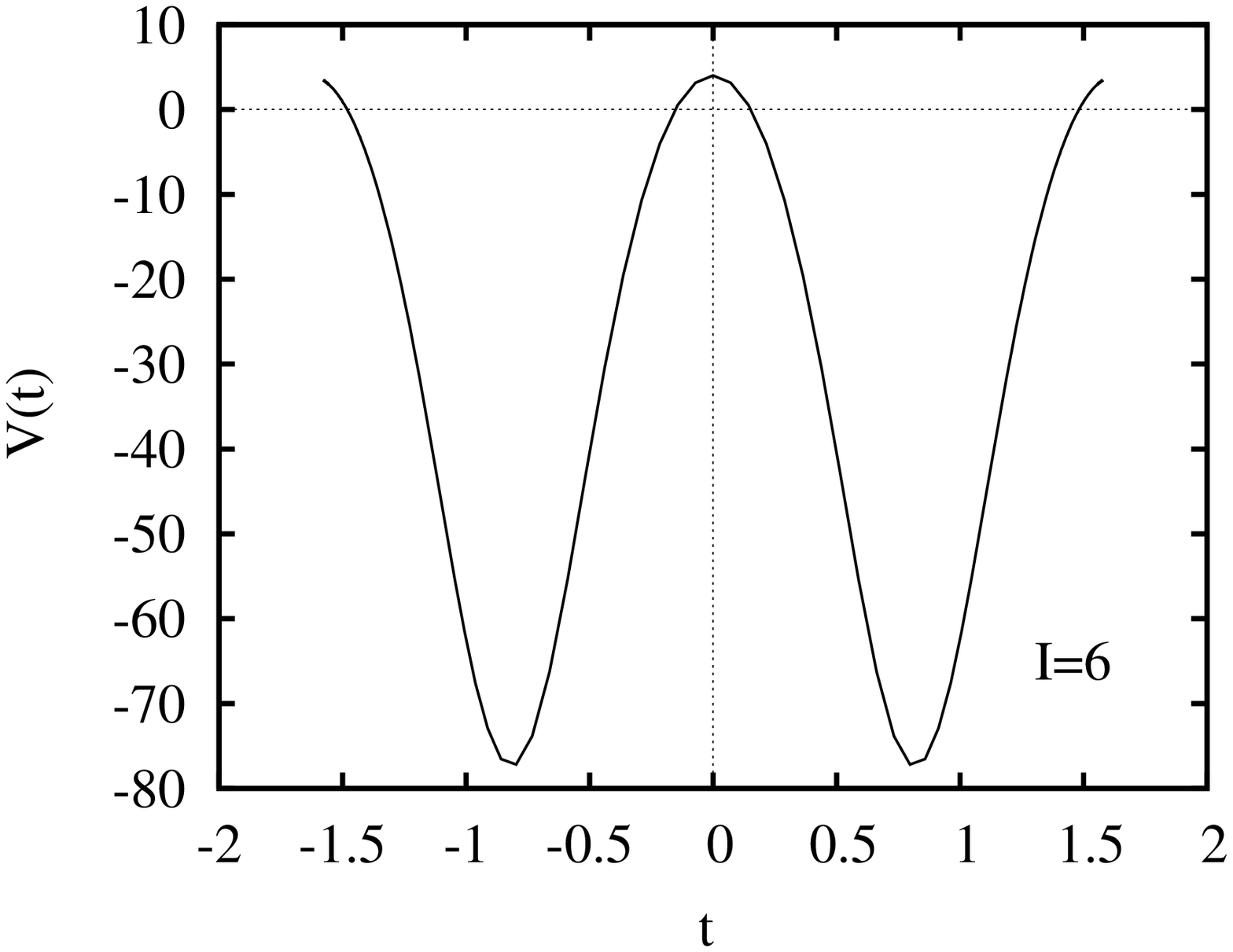}\includegraphics[width=0.5\textwidth]{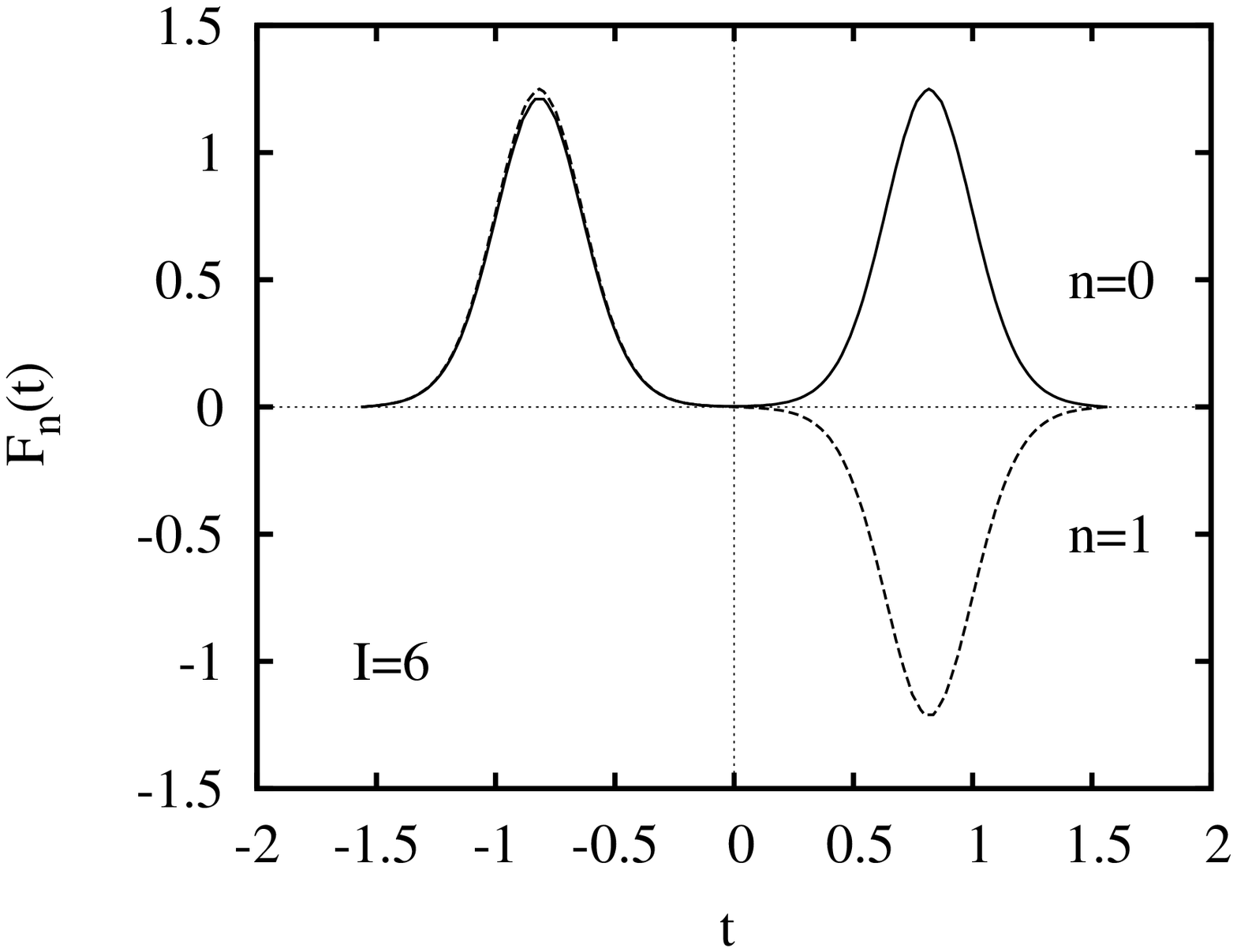}
\begin{minipage}{7cm}
\caption{The dependence of the potential on the variable t, defined by Eq. (2.38). }
\label{Fig.4}
\end{minipage}\ \ 
\hspace*{0.5cm}
\begin{minipage}{7cm}
\caption{Eigenfunction $F_n$ as function of t, for n=0,1.}
\label{Fig.5}
\end{minipage}
\end{figure}
The connection of the two variable is established by the relation (\ref{tfuncx}) and visualized in figures 2 and 3 for I=6. In Ref.
\cite{RaBuRa}, the potential energy was considered as function of $x$ , while here its dependence on the variable $t$ is represented in Fig. 3, also for I=6. If one calculates the average of $R_1$ with trial function $|\psi(z)\rangle$  and the result is considered in the two potential minima one obtains that $\langle \hat{R}_1\rangle =\pm I.$ This shows that in one minimum the system rotates around the axis OX, while in the other minimum the rotation is performed around -OX. An useful insight to the system behavior, for a given solution of the Schr\"{o}dinger equation, is obtained by plotting the wavefunction $F_n(t)$ for n=0,1; 3,4 and 5,6 in Figs. 5,6,7 respectively, for I=6. The pair of states represented in each of the mentioned figures are degenerate.
\begin{figure}[t!]
\includegraphics[width=0.5\textwidth]{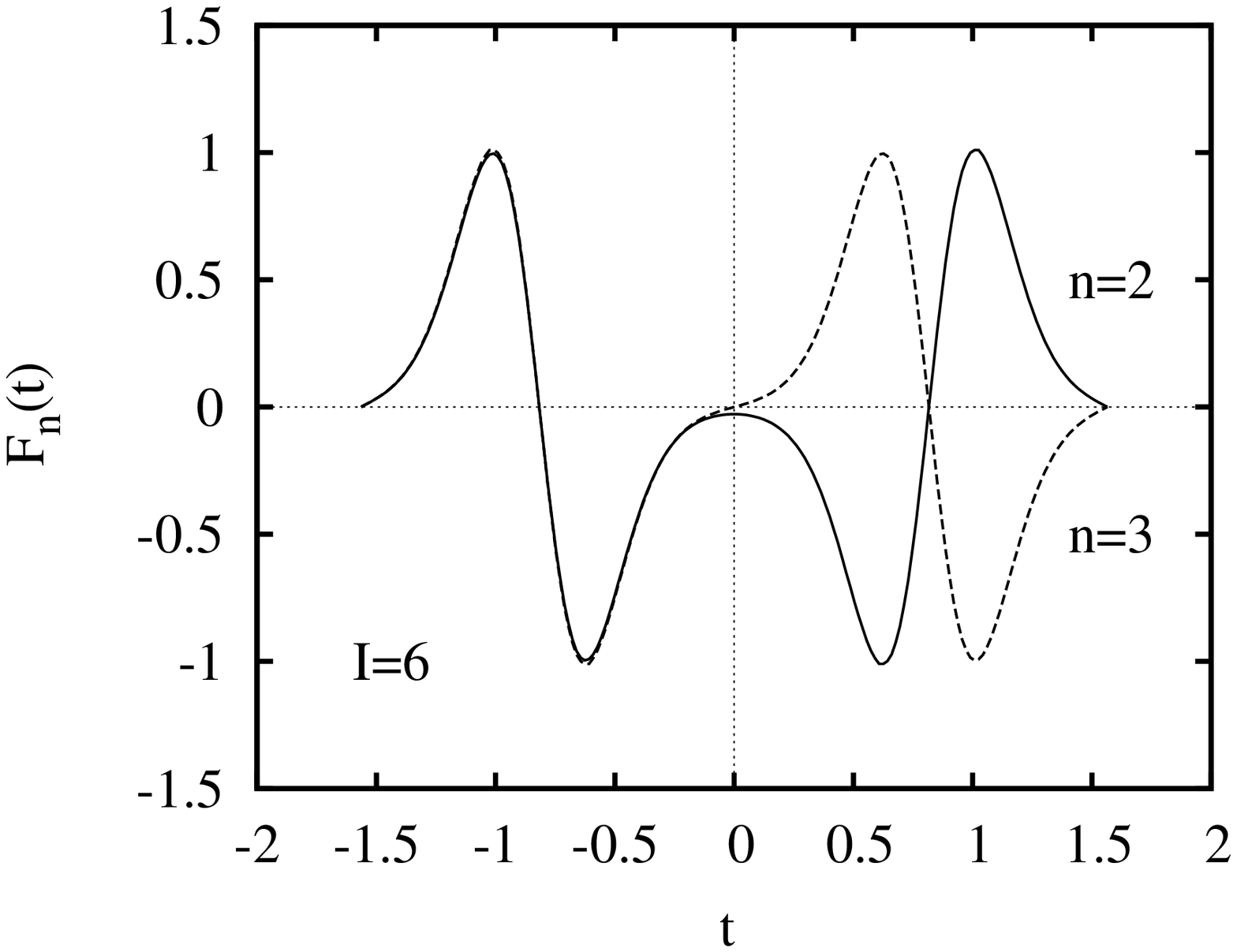}\includegraphics[width=0.5\textwidth]{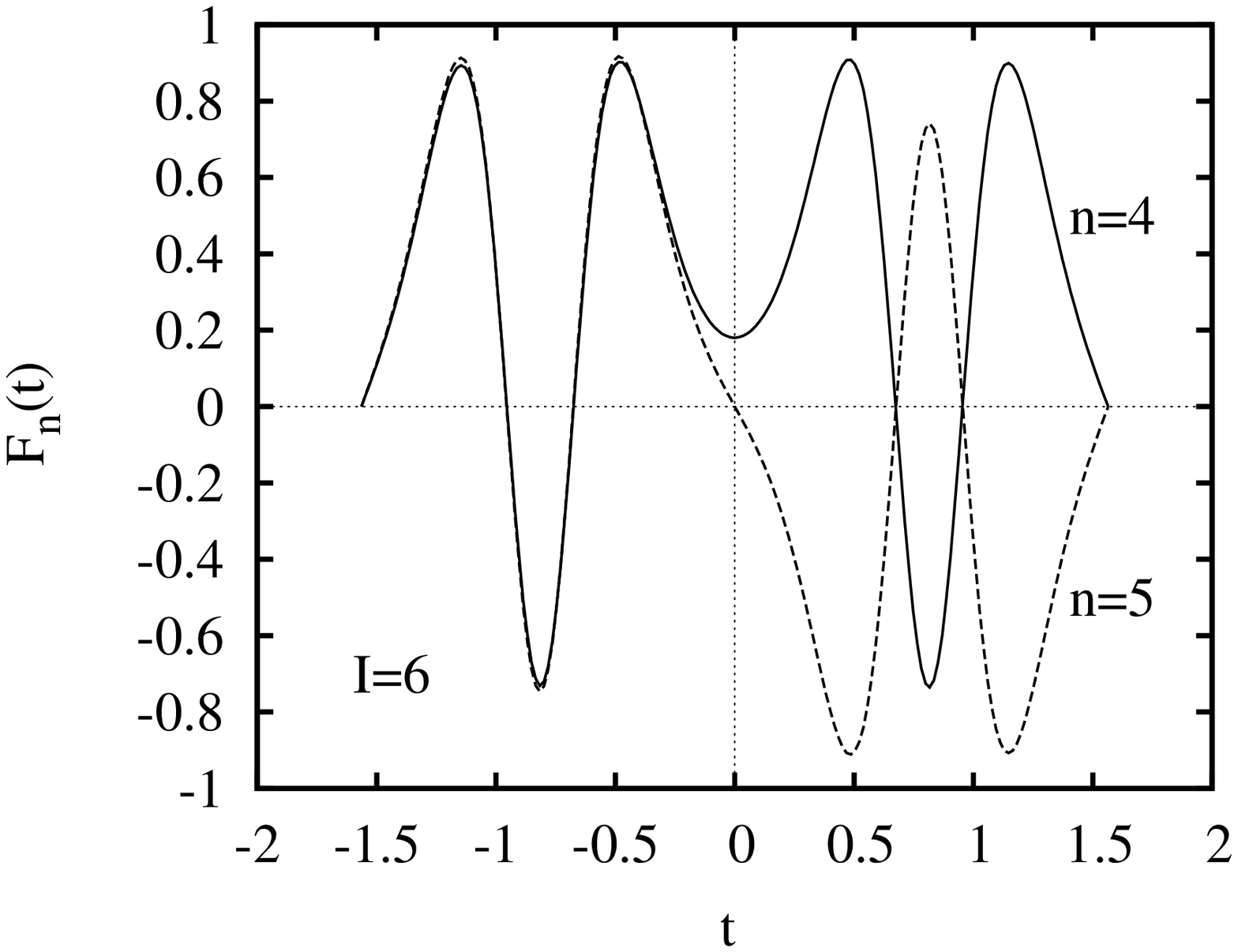}
\begin{minipage}{7.cm}
\caption{The $F_n$ as function of t for n=2,3. }
\label{Fig. 6}
\end{minipage}\ \
\hspace*{0.5cm}
\begin{minipage}{7.cm}
\caption{The $F_n$ as function of t for n=4,5. }
\label{Fig.7}
\end{minipage}
\end{figure}
The probability distributions $|F_n|^2$ for the degenerate states  are identical. Note that if the states corresponding to $F_n$ and $F_{n+1}$ are degenerate, then the states described by 
$F_n+F_{n+1}$ and $F_n-F_{n+1}$ are also degenerate and localized each in a separate well.  For an I running from zero to $I_{max}$,  the lowest two eigenstates of the Schr\"{o}dinger equation for each I form two degenerate bands, one localized inside the well corresponding to the positive minimum and one in the well associated to the negative minimum. The same is also true for the next two degenerate bands and so on.
\begin{figure}[ht!]
\begin{center}
\includegraphics[width=0.7\textwidth]{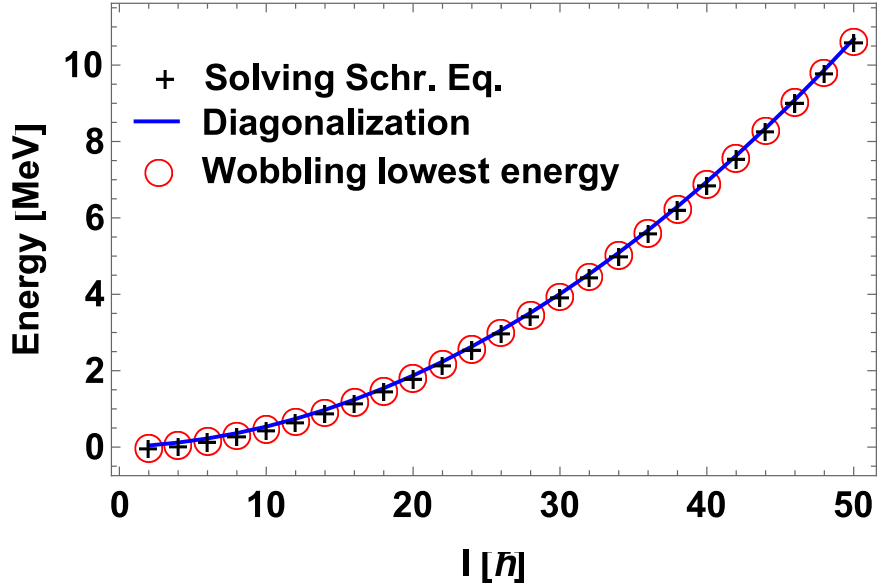}
\end{center}
\caption{Comparison of the lowest exact eigenvalues obtained by diagonalization of $H_R$, the yrast solutions of the Schrodinger equation and the yrast wobbling energies. }
\label{Fig.8}
\end{figure}
In Fig. 8 we compare the yrast energies provided by the three methods: diagonalization, solving the Schr\"{o}dinger equation and by the wobbling energy formula (2.43). We notice that the three sets of energies are almost equal to each other the maximal deviations being less that 5 keV. It is known that for triaxial nuclei, i.e. when the three moments of inertia are all different, the projection of ${\bf R}$ on the OZ axis is not a good quantum number. The present result shows that the coherent states we used as trial function is an optimal mixture of the K components to approximate the exact wavefunction. Also the wobbling approximation of the yrast energies describes very well the exact solution of the Schr\"{o}dinger equation.

The diagonalization procedure provides the wave functions corresponding to the (2J+1) eigenvalues:
\begin{equation}
\Psi^{diag;i}_{JM}=\sum_{K=-J}^{+J} A^{diag;i}_{JK}|JMK\rangle,\;\;i=1,2,...,(2J+1)
\end{equation}
Among the (2J+1) eigenvalues one identifies J degenerate doublets and one non-degenerate level. The mentioned degeneracy is caused by the D2 symmetry satisfied by the rotor Hamiltonian.
It is interesting to see how the quantum number corresponding to the maximum amplitude varies with $i$ for a given angular momentum J. This dependence is given in Fig. 9. Again, we notice that for $i=1$ the maximal component has $K=0$.
\begin{figure}[ht!]
\begin{center}
\includegraphics[width=0.9\textwidth]{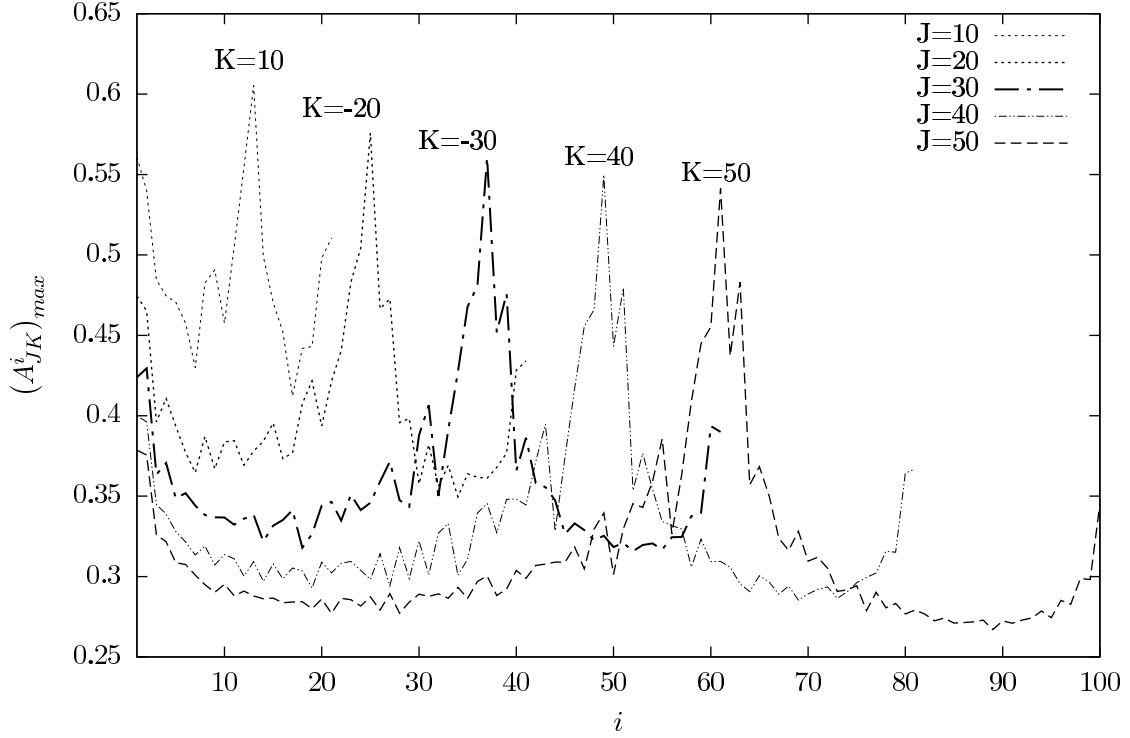}
\end{center}
\caption{The maximal K- amplitude for the i-th exact eigenfunction for a given I.}
\label{Fig.9}
\end{figure}
The comparison of the exact eigenvalues obtained through diagonalization and those obtained by solving the Schr\"{o}dinger equation is achieved in Figs. 10 and 11 for the first six excitation energies as well as for the next five. We notice that the two sets of energies almost coincide with each other. However, this does not happen when the Schr\"{o}dinger equation solutions are compared with the corresponding wobbling energies as shown in Figs 12 and 13 for the i-th  solutions with i=3,4...,10.
\begin{figure}[ht!]
\begin{center}
\includegraphics[width=0.6\textwidth]{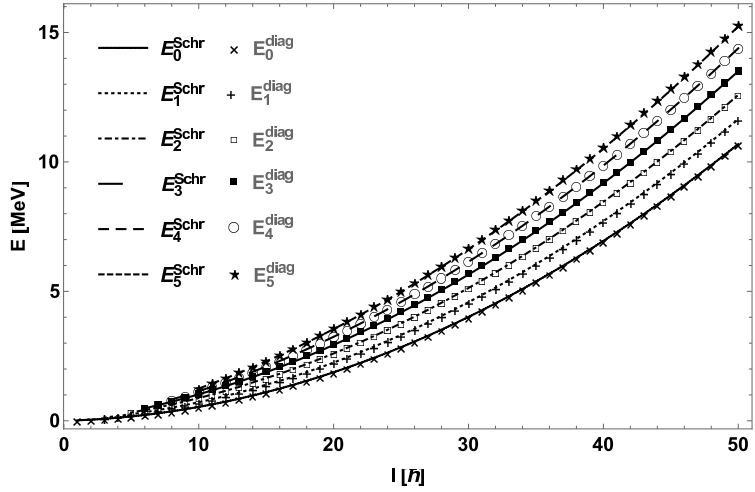}
\end{center}
\caption{The lowest sixth solutions for a given angular momentum I, given by diagonalizing $H_R$ and by solving the Schrodinger equation, respectively.}
\label{Fig.10}
\end{figure}
\begin{figure}[ht!]
\begin{center}
\includegraphics[width=0.6\textwidth]{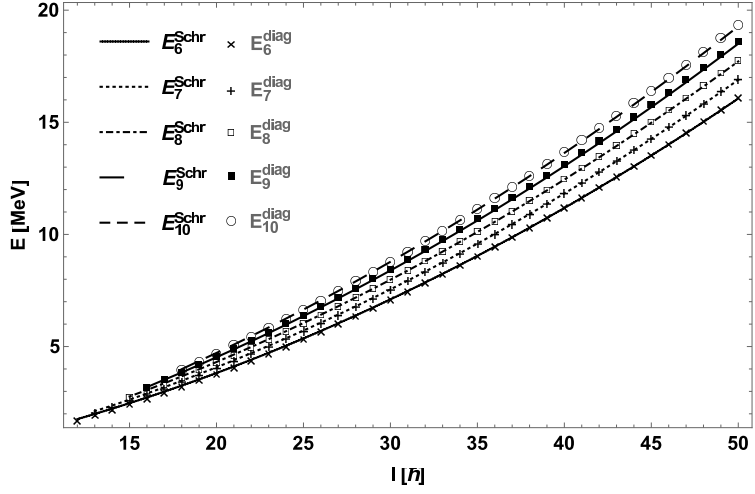}
\end{center}
\caption{The  7-th up to 11-th solutions for a given angular momentum I, given by diagonalizing $H_R$ and by solving the Schrodinger equation, respectively.}
\label{Fig.11}
\end{figure}
\begin{figure}[ht!]
\begin{center}
\includegraphics[width=0.6\textwidth]{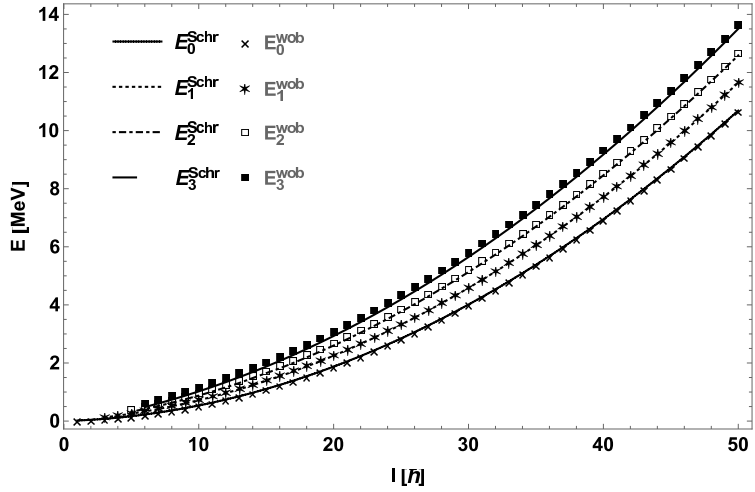}
\end{center}
\caption{The first four solutions for the Schrodinger equations and the first four wobbling energies, respectively.}
\label{Fig.12}
\end{figure}
\begin{figure}[ht!]
\begin{center}
\includegraphics[width=0.6\textwidth]{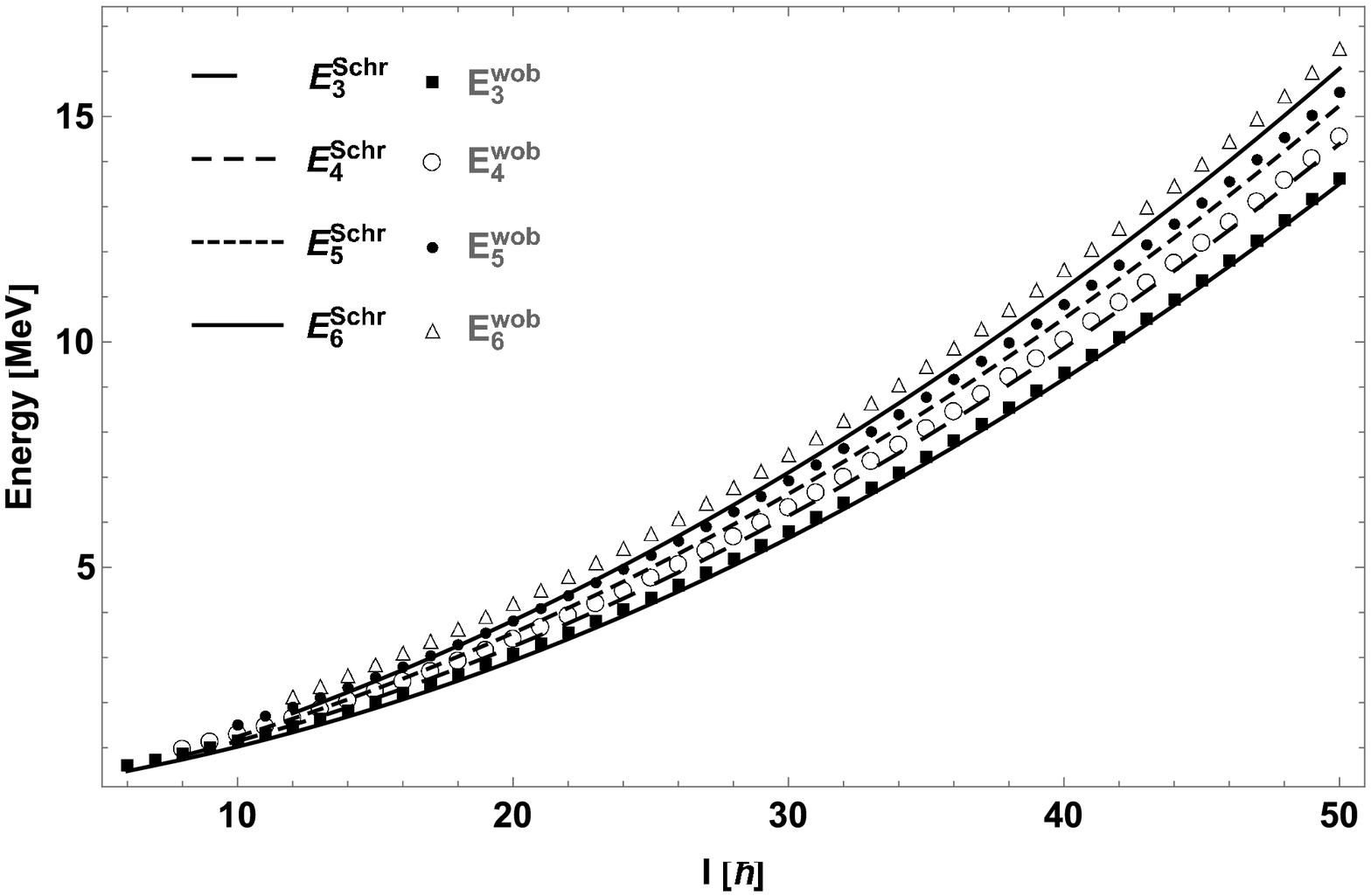}
\end{center}
\caption{The fourth up to eighth  solutions for the Schrodinger equations and the fourth up to seventh wobbling energies, respectively.}
\label{Fig.13}
\end{figure} 
Of course, we may ask ourself how the  harmonic approximation of the Bargmann representation 
(\ref{yra}) with fitted moment of inertia, describes the experimental data for $^{158}$Er.  The agreement quality is shown in Fig.\ref{yrer}, where the experimental and calculated energies are represented as function of I. Theoretical results were obtained with the wobbling formula (2.43). We notice that the wobbling frequency depends almost linearly on the angular momentum I. This dependence can be seen in Fig.\ref{wober}. Details about fixing the moments of inertia are as follows. The expression for the yrast energies can be put in a more suitable form:
\begin{equation}
E^{yr}_{J}=aJ(J+1)+b\sqrt{J(J+1)},
\label{yra1}
\end{equation}
with the evident notations:
\begin{equation}
a=\frac{1}{2{\cal J}_1},\;\;b=\frac{1}{2}\sqrt{(\frac{1}{{\cal J}_2}-\frac{1}{{\cal J}_1})
(\frac{1}{{\cal J}_3}-\frac{1}{{\cal J}_1})}.
\label{abexpr}
\end{equation}
The parameters $a$ and $b$ were fixed by fitting the experimental data taken from Ref. \cite{Nica}
for J=even, with the calculated energies from Eq. (\ref{yra1}). The used fitting procedure is that of the least mean square. The result for parameters $a$ and $b$ is:
\vskip0.2cm
$$a=0.005087581 MeV;\;\;b=0.138354551MeV$$
\vskip0.2cm
The first equation (\ref{abexpr}) leads to ${\cal J}_1=98.278533551[\hbar^2MeV^{-1}]$ , while the second  one provides a relation connecting ${\cal J}_2$ and ${\cal J}_3$.
With these parameters, Eq.(\ref{yra1}) gives the energies for yrast even-spin levels, which are represented in Fig.\ref{yrer}  as a function of the angular momentum and compared with the corresponding experimental data. Also the wobbling frequency is shown in Fig. \ref{wober} as function of I. One notice that the wobbling frequency depends almost linearly on the angular momentum I.     
\begin{figure}
\includegraphics[width=0.5\textwidth]{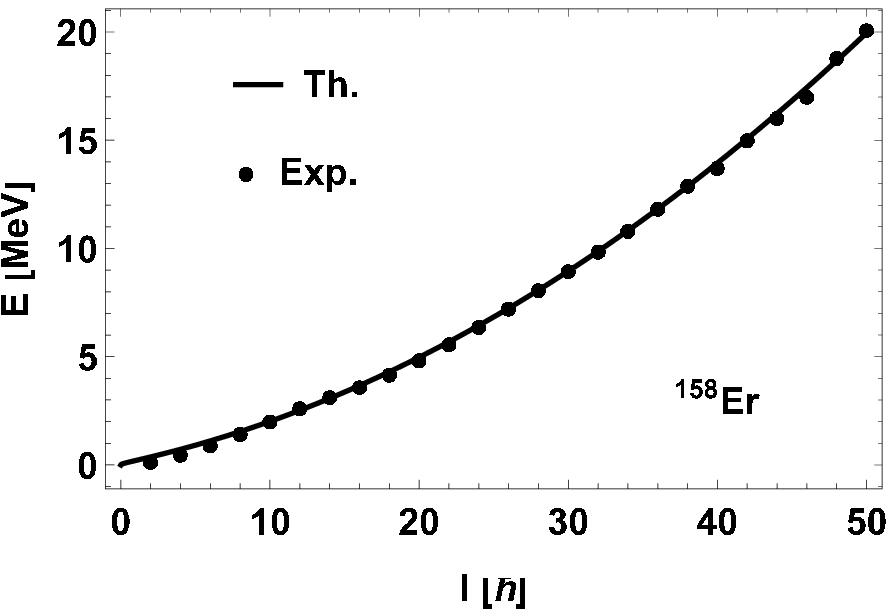}\includegraphics[width=0.5\textwidth]{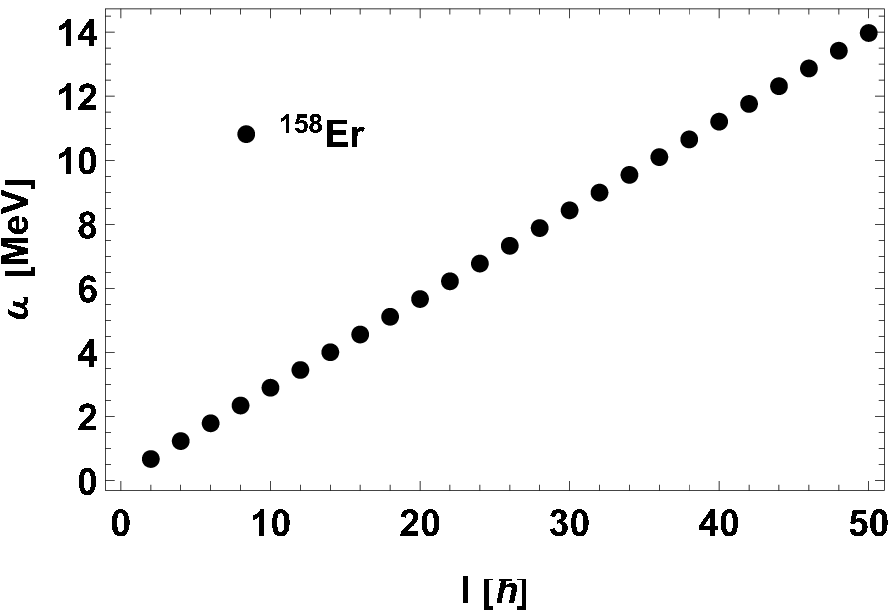}
\begin{minipage}{7cm}
\caption{Results (Th.) obtained with Eq.(2.42) are compared with experimental data (Ex.) taken from \cite{Nica}.}
\label{yrer}
\end{minipage}\ \ \hspace*{0.5cm}
\begin{minipage}{7cm}
\caption{The wobbling frequency for $^{158}$Er as function of the angular momentum.}
\label{wober}
\end{minipage}
\end{figure} 
\begin{table}
\begin{tabular}{|c|cc|c|c|c|c|}
\hline
&\multicolumn{2}{c|}{$B(E2;In_w\to (I-2)n^{'}_w)$} &$B(E2;(I+1)n_w\to In^{'}_w)$&
$B(M1;(I+1)n_w\to In^{'}_w)$&$Q_{I+1}$ & $\mu_{I+1}$\\
&\multicolumn{2}{c|}{$[W.u]$}    &  $[W.u]$&$[10^{-4}\mu_N^2]$&$[e.fm^2]$&$[\mu_N]$\\
\hline
&\multicolumn{2}{c|}{$n_w=n^{'}_w=0$}  & $n_w=1\;n^{'}_w=0$&$n_w=1\;n^{'}_w=0 $ &    &             \\
I& Th.         &       Exp. \cite{Nica}    & Th. &Th. &Th.&  Th.          \\
\hline
2&   67.211    &      129$\pm $9         &0.697&0.143&-79.868&1.291 \\       
4&  186.697    &      186$\pm$ 6         &0.368&0.346&-109.756&2.151   \\
6&  232.653    &      246$\pm$ 8         &0.223&0.473&-127.295&3.012        \\
8&  256.983    &      298$\pm$ 10        &0.149&0.538&-138.606&3.443  \\
10& 272.044    &      250$\pm$ 4         &0.107&0.573&-146.488& 4.734  \\
12& 282.285    &      260$\pm$3          &0.080&0.595&-152.289&5.594   \\
14& 289.702    &                         &0.062&0.611&-156.735&6.456\\
16& 295.321    &                         &0.050&0.623&-160.250&7.316\\
18& 299.724    &                         &0.041&0.632&-163.099&8.177\\
20& 303.269    &                         &0.034&0.640&-165.454&9.038\\
22& 306.183    &                         &0.029&0.646&-167.434&9.898\\
24& 308.621    &                         &0.024&0.652&-169.121&10.759\\
26& 310.692    &                         &0.021&0.657&-170.576&11.620\\
28& 312.472    &                         &0.018&0.661&-171.843&12.481\\
30& 314.018    &                         &0.016&0.664&-172.957&13.342\\
\hline
\end{tabular}
\caption{The calculated intra-band B(E2) values are compared with the available experimental data, taken from Ref.\cite{Nica}.Also, the calculated B(E2) and B(M1) values connecting the $\Delta I=1$ yrast states as well the theoretical values for the quadrupole and magnetic moments are listed.}
\end{table}
In order to calculate the transition probabilities we need to fix the deformations $\beta$ and 
$\gamma$. 
For the case of $^{158}$Er the nuclear quadrupole deformation is taken equal to 0.203 \cite{Lala}, while $\gamma$ is considered to be a free parameter. Within a more consistent formalism, $\gamma$ would be a dynamical variable. However in our formalism the model states for the wobbling levels do not depend on $\gamma$, which suggests to take for $\gamma$ a rigid value, which hereafter will be denoted by $\gamma_0$. Actually its constant value was fixed so that the experimental 
B(E2) value for a particular transition is reproduced. Thus, we arrived at $\gamma_0=12^{0}$.

Having a rigid character, $\gamma$ is spin independent. 
Such a behavior is at variance with the microscopic description which studied the shape dependence on angular momentum \cite{Lee,Riley,Beck}. For example Ref.\cite{Beck} pointed out that in $^{158}$Er, approaching the band termination the states collectivity decreases which might be caused by the fact that the corresponding shape is almost oblate. Also, the coupling with two or four quasiparticle states may lead to discontinuities in the yrast spectrum 
\cite{Lee,RadBu}. Combination of the single particle level crossing effect and the prolate-oblate competition was accounted for in Ref. \cite{Riley} and a prolate-oblate shape coexistence was pointed out for high spins. Similar result was obtained in Ref.\cite{Shi} when the rotation axis coincides with one principal axis. However, when the rotational axis changes the direction the higher energy minimum becomes a saddle point. As shown in Ref.\cite{Cast} the soft gamma shape is a favored shape at low spins,  while in  the high-spin region the alignment induced by rotation brings the system to an oblate shape. In the transitional region a triaxial shape shows up. The authors arrived at this conclusion by analyzing both the even-odd staggering of the energy levels
and the quadrupole transition probabilities. Although the proposed phenomenological formalism is very simple, it accounts for the main features of the energy spectrum and B(E2) values. Indeed, if in our minimum $\chi^2$ calculations for energies we considered  moments of inertia as given by the hydrodynamical model 
\begin{equation}
{\cal J}_k=\frac{4}{3}{\cal J}_0\sin^2(\gamma-\frac{2\pi}{3}k),\;\;k=1,2,3,
\end{equation}
the variational parameters would be ${\cal J}_0$ and $\gamma$. One finds out that $\chi^2$ has two flat minima, $\gamma_1=0 $ and $\gamma_2=\pi$. These are critical points for the transitions from an axial symmetric to a non-axial symmetric shape and from a prolate to an oblate shape, respectively. In the transitional region the triaxial shapes are all along present. Also, as already mentioned the available data for transition probabilities are reasonable well described by a small $\gamma_0$.

Results for transition probabilities and moments are obtained with an effective charge $e_{eff}=1.4 e$ and collected in Table I. The agreement with the corresponding experimental data for the 
B(E2) values is good.

\section{Description of the wobbling motion in even-odd nuclei}  
\renewcommand{\theequation}{3.\arabic{equation}}
\setcounter{equation}{0}
We suppose that the odd-mass nuclear system consists of an even-even core described by a triaxial rotor Hamiltonian and a single j-shell particle moving in a quadrupole deformed mean-field:
\begin{equation}
H_{sp}=\frac{V}{j(j+1)}\left[\cos\gamma(3j_3^2-{\bf j}^2)-\sqrt{3}\sin\gamma(j_1^2-j_2^2)\right].
\end{equation}
It is convenient to express the rotor Hamiltonian in terms of the total angular momentum ${\bf I}$ and the angular momentum carried by the odd particle:
\begin{equation}
H_{rot}=\sum_{k=1,2,3}A_k(I_k-j_k)^2.
\end{equation}
Where $A_k$ are expressed in terms of the moments of inertia associated to the principal axes of the inertia ellipsoid as:
\begin{equation}
A_k=\frac{1}{2{\cal I}_k}.
\end{equation}
In what follows, the moments of inertia are taken as given by the rigid-body model in the Lund convention:
\begin{equation}
{\cal I}^{rig}_{k}=\frac{{\cal I}_0}{1+(\frac{5}{16\pi})^{1/2}\beta}\left[1-\left(\frac{5}{4\pi}\right)^{1/2}\beta\cos\left(\gamma+\frac{2}{3}\pi k\right)\right],\;k=1,2,3
\end{equation}
To the total Hamiltonian
\begin{equation}
H=H_{rot}+H_{sp}
\end{equation}
we associate the time dependent variational equation
\begin{equation}
\delta\int_{0}^{t}\langle \Psi|H-i\frac{\partial}{\partial t'}|\Psi\rangle d t'=0,
\end{equation}
where the trial function is chosen as:
\begin{equation}
|\Psi\rangle ={\bf N}e^{z\hat{I}_-}e^{s\hat{j}_-}|IMK\rangle |jj\rangle,
\end{equation} 
with $\hat{I}_-$ and $\hat{j}_-$ denoting the lowering operators for the intrinsic angular momenta ${\bf I}$ and ${\bf j}$ respectively, while ${\bf N}$ is  the normalization factor having the expression:
\begin{equation}
{\bf N}^{-2}=(1+|z|^2)^{2I}(1+|s|^2)^{2j}.
\end{equation}
The variables $z$ and $s$ are complex functions of time and play the role of classical phase space coordinates describing the motion of the core and the odd particle, respectively:
\begin{equation}
z=\rho e^{i\varphi},\;\;s=fe^{i\psi}
\end{equation}
The variables $(\varphi, r)$ and $(\psi,t)$ with $r$ and $t$ defined as:
\begin{eqnarray}
r&=&\frac{2I}{1+\rho^2},\;\;0\le r\le 2I\nonumber\\ 
t&=&\frac{2j}{1+f^2},\;\; 0\le t\le 2j,
\end{eqnarray}
bring the classical equations, provided by the variational principle, to the canonical form:
\begin{eqnarray}
\frac{\partial {\cal H}}{\partial r}&=&\stackrel{\bullet}{\varphi};\;\frac{\partial {\cal H}}{\partial \varphi}=-\stackrel{\bullet}{r} \nonumber\\ 
\frac{\partial {\cal H}}{\partial t}&=&\stackrel{\bullet}{\psi};\;\frac{\partial {\cal H}}{\partial \psi}=-\stackrel{\bullet}{t}. 
\label{eqmot}
\end{eqnarray}
where ${\cal H}$ denotes the average of $H$ (Eq.3.5) with the function $|\Psi\rangle$ and has the expression:
\begin{eqnarray}
{\cal H}&=&\frac{I}{2}(A_1+A_2)+A_3I^2+\frac{2I-1}{2I}r(2I-r)\left(A_1\cos^2\varphi+A_2\sin^2\varphi -A_3\right)\nonumber\\
        &+&\frac{j}{2}(A_1+A_2)+A_3j^2+\frac{2j-1}{2j}t(2j-t)\left(A_1\cos^2\psi+A_2\sin^2\psi -A_3\right)\nonumber\\
        &-&\sqrt{r(2I-r)t(2j-t)}\left(A_1\cos\varphi\cos\psi+A_2\sin\varphi\sin\psi\right)+A_3\left(r(2j-t)+t(2I-r)\right)\nonumber\\
        &+&V\frac{2j-1}{j+1}\left[\cos\gamma-\frac{t(2j-t)}{2j^2}\sqrt{3}\left(\sqrt{3}\cos\gamma -\sin\gamma\cos2\psi\right)\right].
\end{eqnarray}
From Eq.(\ref{eqmot}) we see that the angles $\varphi$ and $\psi$ play the role of generalized coordinates while $r$ and $t$  are the corresponding conjugate momenta.
Looking for the extremal points of the energy surface ${\cal H}=const$, one finds out that the point $(\varphi,r;\psi,t)=(0,I;0,j)$ is a minimum point for the classical energy function.
Aiming at a compact expression for the equations to be used in what follows, it is convenient to introduce the notations:
\begin{equation}
q_1=\varphi,\;\;q_2=\psi,\;\;p_1=r,\;\;p_2=t.
\end{equation}
Performing a linear expansion in the left hand side of Eq.(\ref{eqmot}), around the mentioned minimum point of ${\cal H}$, one finds:
\begin{eqnarray}
\stackrel{\bullet}{q}^{\prime}_i&=& \sum_{k=1,2}A_{ik}p^{\prime}_k,\nonumber\\
\stackrel{\bullet}{p}^{\prime}_i&=& \sum_{k=1,2}B_{ik}q^{\prime}_k,
\label{eqqandp}
\end{eqnarray}
where the deviation of the current variables from the corresponding minimum value is denoted with a similar symbol but accompanied by $\prime$.
The matrices A and B are given explicitly in Appendix A. 

It is useful to express the equations (\ref{eqqandp}) in terms of the complex coordinates:
\begin{equation}
a_k=\frac{q_k+ip_k}{\sqrt{2}},\;\;a^{*}_k=\frac{q_k-ip_k}{\sqrt{2}}.
\end{equation}
Further, we determine the complex variable
\begin{equation}
C^*=\sum_{k=1,2}\left(R_ka^*_k-S_ka_k\right),
\end{equation}
such that the following equations are fulfilled:
\begin{equation}
\{C^*,{\cal H}\}=i\omega C^*,\;\; \{C^*,C\}=i
\label{Poissonbr}
\end{equation}
with $\{,\}$ denoting the Poisson bracket.
The first  equation (3.17) leads to a homogeneous system of linear equations of random phase approximation (RPA) type:
\begin{eqnarray}
\left(\begin{matrix}\frac{A-B}{2} & -\frac{A+B}{2}\cr
                     \frac{A+B}{2}&-\frac{A-B}{2} \end{matrix}\right)\left(\begin{matrix}R&\cr S&\end{matrix}\right)=\Omega \left(\begin{matrix}R&\cr S&\end{matrix}\right).
\end{eqnarray}
These equations determine the amplitudes $R$ and $S$ up to a multiplicative constant which is fixed by the second relation of (3.17).
The compatibility condition for this system of homogeneous equations yields the equation for $\Omega$:
\begin{equation}
\Omega^4+B\Omega^2+C=0.
\label{Ome}
\end{equation}
where the coefficients B and C are given by
\begin{eqnarray}
-B&=&\left[(2I-1)(A_3-A_1)+jA_1\right]\left[(2I-1)(A_2-A_1)+jA_1\right]+2A_2A_3Ij\nonumber\\
 &+&\left[(2j-1)(A_3-A_1)+IA_1+V\frac{2j-1}{j(j+1)}\sqrt{3}(\sqrt{3}\cos\gamma+\sin\gamma)\right]\nonumber\\
 &\times&\left[(2j-1)(A_2-A_1)+IA_1+V\frac{2j-1}{j(j+1)}2\sqrt{3}\sin\gamma\right],\\
C&=&\left\{\left[(2I-1)(A_3-A_1)+jA_1\right] \left[(2j-1)(A_3-A_1)+IA_1+V\frac{2j-1}{j(j+1)}\sqrt{3}(\sqrt{3}\cos\gamma+\sin\gamma)\right]\right.\nonumber\\
&&\left.-IjA_3^2\right\}\nonumber\\
 &\times&\left\{\left[(2I-1)(A_2-A_1)+jA_1\right]\left[(2j-1)(A_2-A_1)+IA_1+V\frac{2j-1}{j(j+1)}2\sqrt{3}\sin\gamma\right]-IjA_2^2\right\}.\nonumber\\
\end{eqnarray}
There exists a certain interval for the parameters to be fixed, where Eq.(\ref{Ome}) admits two real and positive solutions:
\begin{eqnarray}
\left(\begin{matrix}\Omega_1\cr\Omega_2\end{matrix}\right)=\left[\frac{1}{2}\left(-B\mp(B^2-4C)^{1/2}\right)\right]^{1/2}.
\end{eqnarray}
Finally, the semiclassical  eigenvalues of H (3.5) are given by:
\begin{eqnarray}
E_{I,n_1,n_2}&=&{\cal H}_{min}(I)+\hbar\Omega_1(n_1+\frac{1}{2})+\hbar\Omega_2(n_2+\frac{1}{2}),\;\rm{with}\\
{\cal H}_{min}(I)&=&\frac{I+j}{2}\left(A_2+A_3\right)+\left(I^2+j^2-Ij\right)A_1+2IjA_3-V\frac{2j-1}{j+1}\sin(\gamma+\frac{\pi}{6}).\nonumber
\end{eqnarray}

\subsection{An alternative description}
Here we present a slightly different method to derive analytical expression for the wobbling frequency of the even-odd system  
, which are easier to be manipulated in the fitting calculations for  the experimental data.
We start by expanding the classical energy function around the minimum point, in the second order of approximation:
\begin{eqnarray}
{\cal H}&=& {\cal H}_{min}+\frac{1}{I}\left[(2I-1)(A_3-A_1)+jA_1\right]\frac{r'^2}{2}-\frac{1}{2}A_3r't'\nonumber\\
        &+&\frac{1}{j}\left[(2j-1)(A_3-A_1)+IA_1+V\frac{2j-1}{j(j+1)}\sqrt{3}(\sqrt{3}\cos\gamma+\sin\gamma)\right]\frac{t'^2}{2}-\frac{1}{2}A_3r't'\nonumber\\
        &+&I\left[(2I-1)(A_2-A_1)+jA_1\right]\frac{\varphi'^2}{2}-\frac{1}{2}IjA_2\varphi'\psi'\nonumber\\
        &+&j\left[(2j-1)(A_2-A_1)+IA_1+V\frac{2j-1}{j(j+1)}2\sqrt{3}\sin\gamma\;\frac{\psi'^2}{2}\right]-\frac{1}{2}IjA_2\varphi'\psi'.
\end{eqnarray}
If one ignores the coupling terms, the remaining Hamiltonian describes two uncoupled oscillators whose frequencies are:
\begin{eqnarray}
\omega_1&=&\left[(2I-1)(A_3-A_1)+jA_1\right]^{1/2}\left[(2I-1)(A_2-A_1)+jA_1\right]^{1/2},\nonumber\\
\omega_2&=&\left[(2j-1)(A_3-A_1)+IA_1+V\frac{2j-1}{j(j+1)}\sqrt{3}\left(\sqrt{3}\cos\gamma+\sin\gamma\right)\right]^{1/2}\nonumber\\
        &\times&\left[(2j-1)(A_2-A_1)+IA_1+V\frac{2j-1}{j(j+1)}2\sqrt{3}\sin\gamma\right]^{1/2}.
\end{eqnarray}
Now, we proceed to treat the coupling terms. To this goal we  quantize the classical coordinates:
\begin{eqnarray}
&&\varphi '\to \hat{q}_1,\;\;r'\to \hat{p}_1,\;\;[{\hat q}_1,\hat{p}_1]=i,\nonumber\\
&&\psi '\to \hat{q}_2,\;\;t'\to \hat{p}_2,\;\;[\hat{q}_2,\hat{p}_2]=i.
\end{eqnarray}
The corresponding creation and annihilation operators are defined by:
\begin{eqnarray}
\hat{q}_1&=&\frac{1}{\sqrt{2}k}\left(a^{\dagger}+a\right),\;\;\hat{p}_1=\frac{ik}{\sqrt{2}}\left(a^{\dagger}-a\right),\nonumber\\
\hat{q}_2&=&\frac{1}{\sqrt{2}k'}\left(b^{\dagger}+b\right),\;\;\hat{p}_2=\frac{ik'}{\sqrt{2}}\left(b^{\dagger}-b\right).
\end{eqnarray}
where the canonicity factors k and k' are chosen such that the uncoupled oscillator Hamiltonian be diagonal, with the result:
\begin{eqnarray}
k&=&\left[\frac{(2I-1)(A_2-A_1)+jA_1}{(2I-1)(A_3-A_1)+jA_1}I^2\right]^{1/4},\nonumber\\
k'&=&\left[\frac{(2j-1)(A_2-A_1)+IA_1+V\frac{2j-1}{j(j+1)}2\sqrt{3}\sin\gamma}{(2j-1)(A_3-A_1)+IA_1+V\frac{2j-1}{j(j+1)}\sqrt{3}\left(\sqrt{3}\cos\gamma+\sin\gamma\right)}j^2\right]^{1/4}.
\end{eqnarray}
The quantized Hamiltonian looks like:
\begin{eqnarray}
H&=&{\cal{H}}_{min}+\hbar\omega_1(a^{\dagger}a+\frac{1}{2})+\hbar\omega_2(b^{\dagger}b+\frac{1}{2})\nonumber\\
 &+&\frac{A_3}{2}kk'\left(a^{\dagger}b^{\dagger}+ba-a^{\dagger}b-b^{\dagger}a\right)-Ij\frac{A_2}{2}\frac{1}{kk'}\left(a^{\dagger}b^{\dagger}+ba+a^{\dagger}b+b^{\dagger}a\right).
\end{eqnarray}
The equations of motion for the creation and annihilation operators are:
\begin{eqnarray}
\left[H,a^{\dagger}\right]&=&\hbar\omega_1a^{\dagger}+\frac{A_3}{2}kk'(b-b^{\dagger})-Ij\frac{A_2}{2}\frac{1}{kk'}(b+b^{\dagger}),\nonumber\\
\left[H,b^{\dagger}\right]&=&\hbar\omega_2b^{\dagger}+\frac{A_3}{2}kk'(a-a^{\dagger})-Ij\frac{A_2}{2}\frac{1}{kk'}(a+a^{\dagger}),\nonumber\\
\left[H,a\right]&=&-\hbar\omega_1a-\frac{A_3}{2}kk'(b^{\dagger}-b)+Ij\frac{A_2}{2}\frac{1}{kk'}(b+b^{\dagger}),\nonumber\\
\left[H,b\right]&=&-\hbar\omega_2b-\frac{A_3}{2}kk'(a^{\dagger}-a)+Ij\frac{A_2}{2}\frac{1}{kk'}(a+a^{\dagger}).
\end{eqnarray}
Now, we define the phonon operator
\begin{equation}
\Gamma^{\dagger}=X_1a^{\dagger}+X_2b^\dagger-Y_1a-Y_2b,
\label{fonen}
\end{equation}
where the amplitudes X and Y are determined such that the following restrictions are fulfilled:
\begin{equation}
\left[H,\Gamma^{\dagger}\right]=\hbar \Omega^{\dagger},\;\;\left[\Gamma,\Gamma^{\dagger}\right]=1.
\label{Gam}
\end{equation}
Taking into account the Dyson boson representation of the angular momenta $I_+$ and $j_+$, one obtains that the bosons $a^{\dagger}$ and $b^{\dagger}$ are tensors of rank 1 and projection 1 with respect to the rotations generated by the angular momenta components $I_k$,k=1,2,3 and 
$j_i$,i=1,2,3, respectively, and consequently so the phonon operator is. Note that Eqs.(\ref{Gam}) are specific to the RPA approach, although the transformation (\ref{fonen}) is, according to the above remark, a Hartree-Fock-Bogoliubov like transformation for the bosons $a^{\dagger}$ and 
$b^{\dagger}$.
The second equation from (\ref{Gam}) leads to:
\begin{equation}
|X_1|^2+|X_2|^2-|Y_1|^2-|Y_2|^2=1,
\end{equation}
while the first one provides a homogeneous system of linear equations for the phonon amplitudes. The compatibility condition can be written either under the form of a dispersion equation:
\begin{eqnarray}
1&=&\frac{1}{\omega_2^2-\omega_1^2}\frac{1}{\omega_1^2-\Omega^2}\left[2A_2A_3Ij\omega_1^2+\left(A_3^2k^2k'^2+I^2j^2\frac{A_2^2}{k^2k'^2}\right)\omega_1\omega_2-A_2^2A_3^2I^2j^2\right]\nonumber\\
 &+&\frac{1}{\omega_2^2-\omega_1^2}\frac{1}{\omega_2^2-\Omega^2}\left[-2A_2A_3Ij\omega_2^2-\left(A_3^2k^2k'^2+I^2j^2\frac{A_2^2}{k^2k'^2}\right)\omega_1\omega_2+A_2^2A_3^2I^2j^2\right]\nonumber\\
&\equiv& f(\Omega).
\label{dispereq}
\end{eqnarray} 
or as an algebraic equation:
\begin{equation}
\Omega^4+B'\Omega^2+C'=0,
\label{ecOm}
\end{equation}
with
\begin{eqnarray}
-B'&=&\omega_1^2+\omega_2^2+2A_2A_3Ij,\nonumber\\
C'&=&\omega_1^2\omega_2^2-\left(A_3^2k^2k'^2+I^2j^2\frac{A_2^2}{k^2k'^2}\right)\omega_1\omega_2+A_2^2A_3^2I^2j^2.
\end{eqnarray}
The energies to be used for describing the experimental data are:
\begin{equation}
E_{I,j,n_1,n_2}={\cal H}_{min}+\hbar\Omega_1(n_1+\frac{1}{2})+\hbar\Omega_2(n_2+\frac{1}{2}).\,\;n_1,\;n_2=0,1,2,....
\label{enerI}
\end{equation}
One may prove analytically that the algebraic equations (3.19) and (\ref{ecOm}) are identical, which is induced by the fact that $B'=B$ and $C'=C$. However, the present form is more suitable for practical purposes. In what follows we shall refer to a given state by mentioning the specific quantum numbers, i.e. $(I,j,n_1,n_2)$. The amplitudes of the phonon operator involved in 
Eq.(\ref{fonen}) are analytically given in Appendic B.
\subsection{Numerical analysis}
\subsubsection{Energies}
The formalism of this Section was used to describe the triaxial super-deformed (TSD) bands TSD1, TSD2, TSD3 and TSD4 in $^{163}$Lu. Experimental excitation energies were taken from 
Ref.\cite{Jens1,Hage}. Since the first three mentioned bands are of positive parity, while the last one of negative parity we assume for the single particle j shells, the states $\pi i_{13/2}$ and $\pi h_{9/2}$, respectively. The choice is suggested by the negative parity orbital which might be occupied by the odd proton, in the spherical shell model. Moreover, as shown in Ref.\cite{Jens1}, a detailed analysis leads to a negative parity assignment for the band TSD4.

\begin{figure}[ht!]
\includegraphics[width=0.5\textwidth]{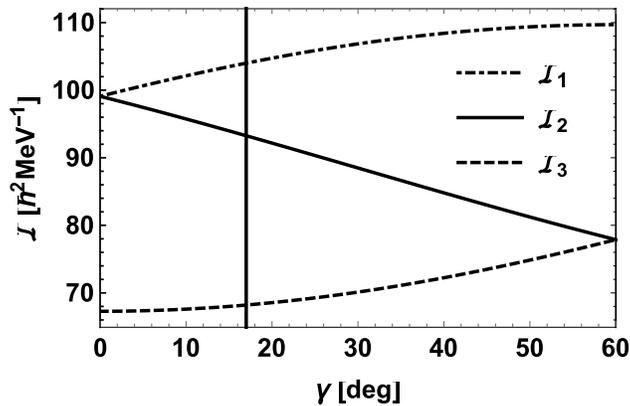}
\caption{The moments of inertia are reprezented as function of $\gamma$. The rigid value of 
$\gamma$, i.e. $\gamma_m=17^{0}$, is also specified by a vertical line.}
\label{inermom}
\end{figure}

It is worth mentioning the fact that we also made the calculations with the option 
$(\pi h_{11/2},2,0)$ for the TSD4, but despite the fact that this configuration would allow a B(E1) transition to the yrast TSD1 band,using a simple expression for the dipole transition operator, the overall agreement with the existent data was poor. This is also one serious reason for which we chose $(\pi h_{9/2},3,0)$ as a basic configuration for TSD4.  
To the bands mentioned above, we shall assign the states $(I,\pi i_{13/2},0,0), (I, \pi i_{13/2}, 1, 0), (I,\pi i_{13/2},2,0), (I,\pi h_{9/2},3,0)$, respectively.
The energies from (\ref{enerI}) depend on two parameters, namely ${\cal I}_0$ and the scaling factor $s (=V{\cal I}_0)$. These were fixed by the least mean square procedure, fitting the mentioned data with our calculated energies for a fixed $\gamma$. The quadrupole nuclear deformation was taken equal to 0.38 \cite{Lala}.  Then we varied $\gamma$ and kept
the value $\gamma_m=17^0$ to which a minimal root mean square for the  deviations of the calculated and experimental energies is obtained. Thus, one obtained the values: $1/{\cal I}_0 =0.0100917\hbar^{-2}MeV$, and $s=6.1937535\hbar^2$. The moments of inertia dependence on the dynamical variable gamma is represented in Fig.\ref{inermom}. Therein the rigid value 
$\gamma_0=17^0$ is also mentioned. 
Using these parameters, we calculated the canonicity factors $k$ and $k'$
whose dependence on the total angular momentum is shown in Figs, 17 and 18 respectively. 
\begin{figure}[ht!]
\hspace*{-1cm}\includegraphics[width=0.5\textwidth]{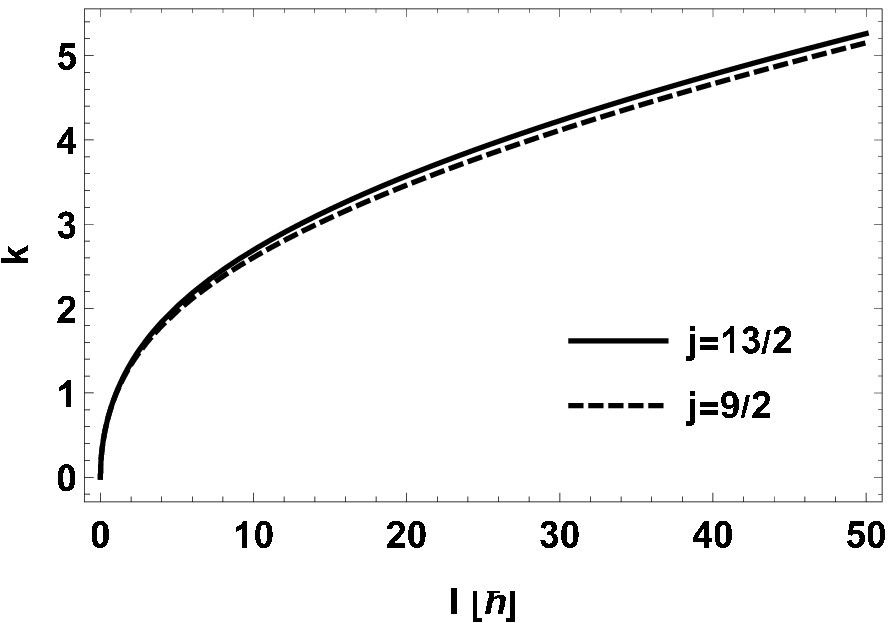}\includegraphics[width=0.5\textwidth]{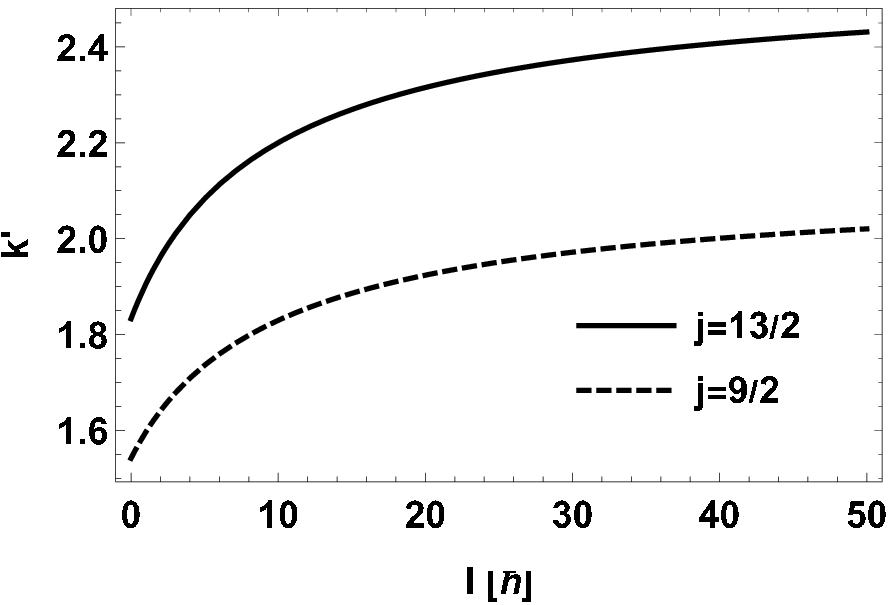}
\begin{minipage}{7cm}
\caption{The canonicity parameter $k_1$ given by Eq.(3.28) as function of the total angular momentum.}
\label{Fig.17}
\end{minipage}\ \
\begin{minipage}{7cm}
\hspace*{0.5cm}
\caption{The canonicity parameter $k_2$ given by Eq.(3.28) as function of the total angular momentum.}
\label{Fig.18}
\end{minipage}
\end{figure}
The phonon energies $\Omega_1$ and $\Omega_2$ depend on I as shown in Figs. 19 and 20, respectively. The dependence of the wobbling energy on the rotation frequency was microscopically studied in Ref.\cite{ShoShi}.
\begin{figure}[ht!]
\hspace*{-1cm}\includegraphics[width=0.5\textwidth]{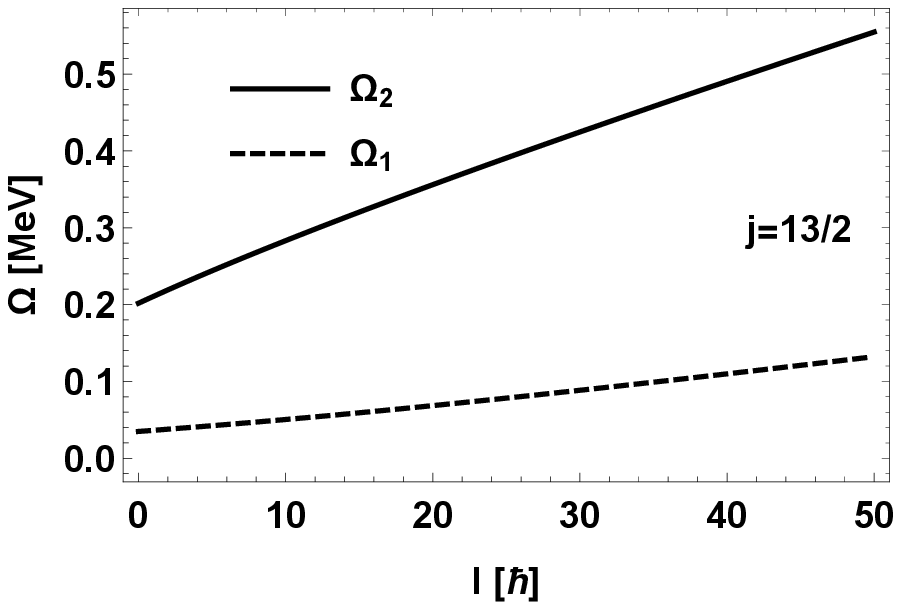}\includegraphics[width=0.5\textwidth]{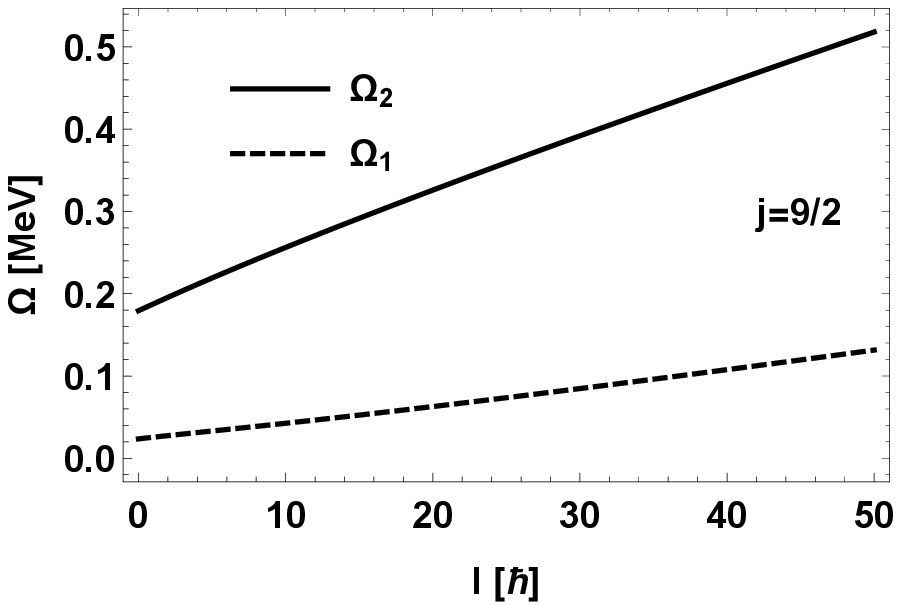}
\begin{minipage}{7cm}
\caption{The phonon energy satisfying Eq.3.35 as function of the total angular momentum I for $j=\pi i13/2$.}
\label{Fig.19}
\end{minipage}\ \
\begin{minipage}{7cm}
\hspace*{0.5cm}
\caption{The two solutions of Eq.(3.35), as function of the total angular momentum, for $j=\pi h_{9/2}$.}
\label{Fig.20}
\end{minipage}
\end{figure}

The calculated energies are compared with the corresponding data in the bands TSD1,TSD2,TSD3 and TSD4 in Figs. 21, 22, 23 and 24, respectively. One remarks the high quality of the agreement with the data.

\begin{figure}[ht!]
\includegraphics[width=0.5\textwidth]{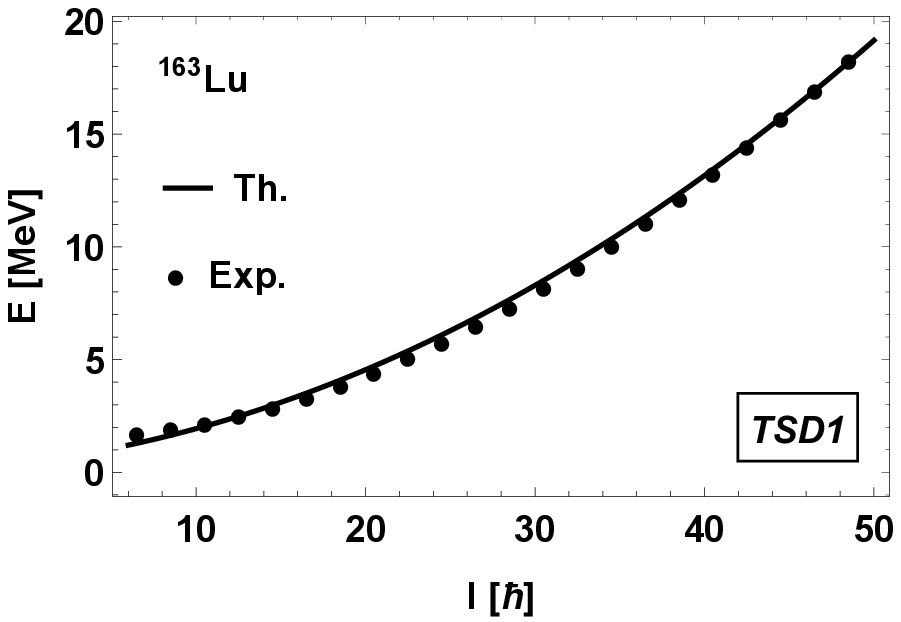}\includegraphics[width=0.5\textwidth]{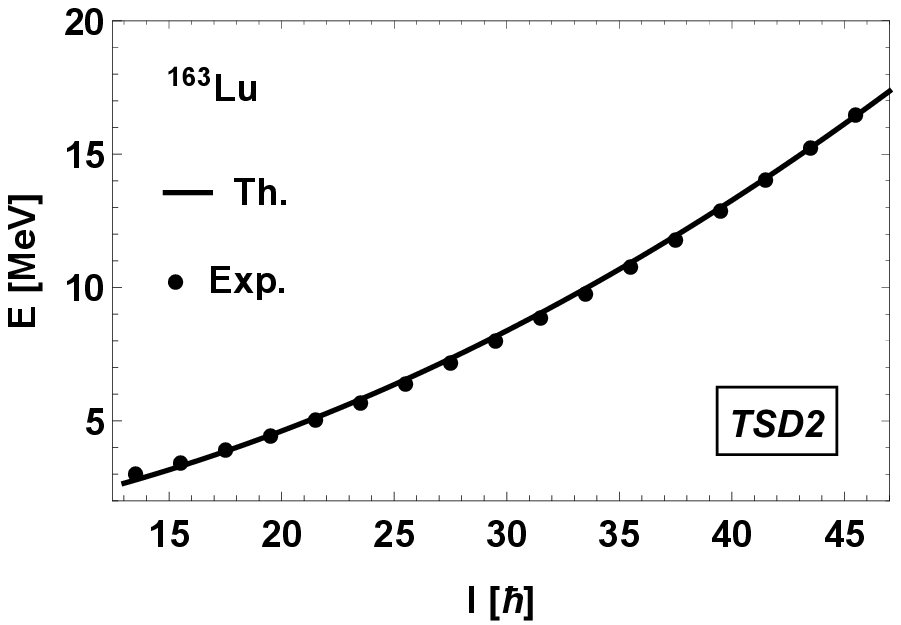}
\begin{minipage}{7cm}
\caption{The excitation energies of the band TSD1. They correspond to the parameters 
$1/{\cal J}_0$ and $s$ obtained by fitting
the experimental corresponding data taken from Ref.\cite{Jens1,Hage}.}
\label{Fig.21}
\end{minipage}\ \
\hspace*{0.5cm}
\begin{minipage}{7cm}
\caption{The excitation energies of the band TSD2. They correspond to the parameters 
$1/{\cal J}_0$ and $s$ obtained by fitting
the experimental corresponding data taken from Ref.\cite{Jens1,Hage}.}
\label{Fig.22}
\end{minipage}
\end{figure}

\begin{figure}[ht!]
\includegraphics[width=0.5\textwidth]{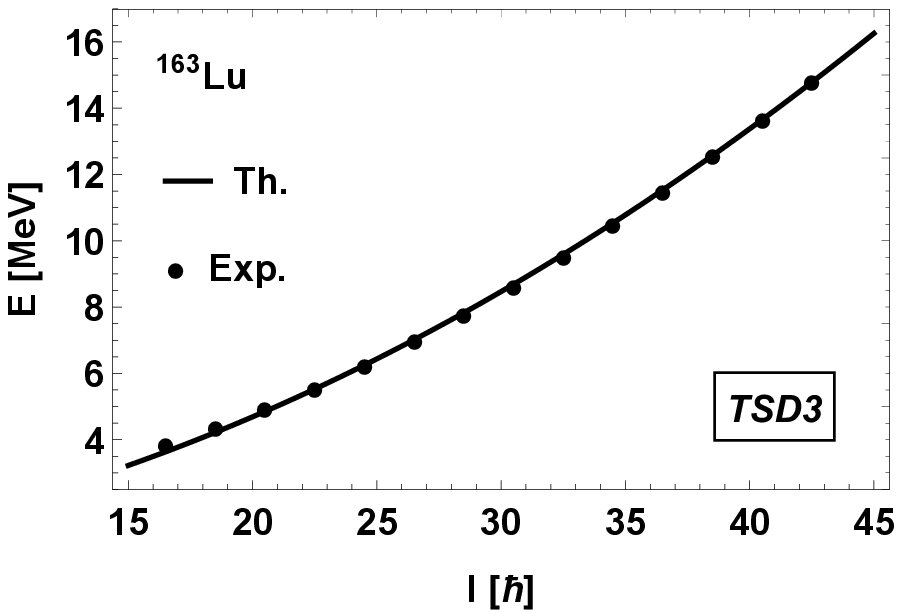}\includegraphics[width=0.5\textwidth]{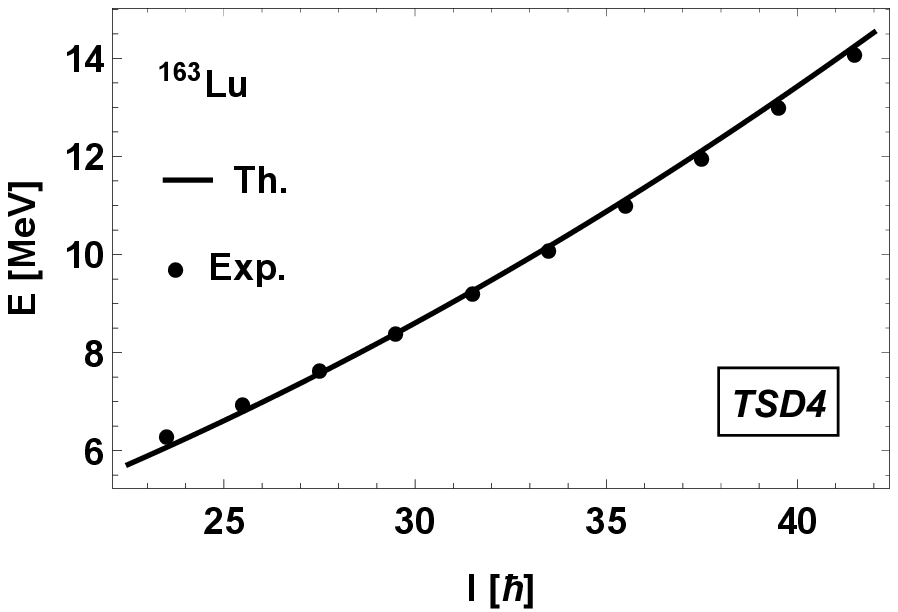}
\begin{minipage}{7cm}
\caption{The excitation energies of the band TSD3. They correspond to the parameters 
$1/{\cal I}_0$ and $s$ obtained by fitting
the experimental corresponding data taken from Ref.\cite{Jens1,Hage}.}
\label{Fig.23}
\end{minipage}\ \
\hspace*{0.5cm}
\begin{minipage}{7cm}
\caption{The excitation energies of the band TSD4. They correspond to the parameters $1/{\cal J}_0$ and $s$ obtained by fitting
the corresponding experimental data taken from Ref.\cite{Jens1,Hage}.}
\label{Fig.24}
\end{minipage}
\end{figure}
In order to see the effect of the coupling terms on the oscillator frequencies $\omega_1$ and $\omega_2$, in Fig. 21 we presented the geometrical solutions of the dispersion equation
(\ref{dispereq}) for I=63/2. The poles have the energy $\omega_1$ and $\omega_2$ with $\omega_1<\omega_2.$. The energies $\Omega_1$ and $\Omega_2$ are obtained by intersecting the curve $f(\Omega)$
with the parallel line to the abscissa axis, of ordinate 1. The first intersection provides $\Omega_1$ while the second one, $\Omega_2$. We see that the coupling diminishes $\omega_1$ and  increases $\omega_2$.
\begin{figure}[ht!]
\begin{center}
\includegraphics[width=0.5\textwidth]{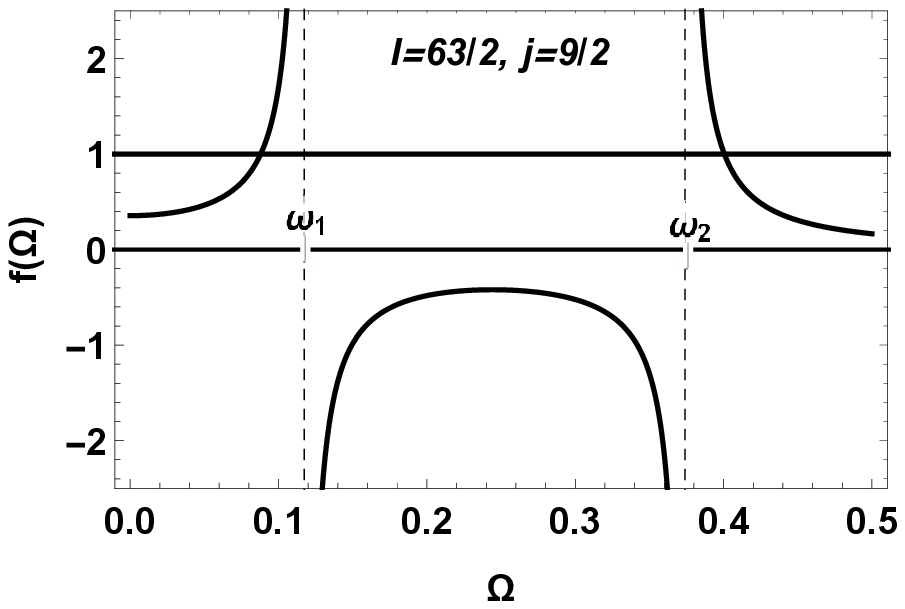}
\end{center}
\caption{The function $f(\Omega)$, defined in Eq.(\ref{dispereq}), is plotted together with the straight line, parallel with the abscissa axis having the ordinate equal to 1. The intersection of the two curves gives $\Omega_1$ and $\Omega_2$. The frequencies for the uncoupled oscillators are just the poles of $f(\Omega)$.}
\label{Fig.25}
\end{figure}
\subsubsection{E.m. transition probabilities}
The  operators used for the E2 and M1 transitions are:
\begin{eqnarray}
{\cal M}(E2,\mu)&=&e_{eff}\left[Q_0D^2_{\mu0}-Q_2(D^2_{\mu2}+D^2_{\mu-2})\right]
+e_{eff}\sum_{\nu}D^2_{\mu\nu}Y_{2\nu}r^2\equiv T^{coll}_{2\mu}+T^{sp}_{2\mu},\nonumber\\
 {\cal M}(M1,\mu)&=&\sqrt{\frac{3}{4\pi}}\mu_N\sum_{\nu=0,\pm1}\left[g_RI_{\nu}+(g_l-g_R)j_{\nu}+(g_s-g_l)s_{\nu}\right]D^1_{\mu\nu}\equiv M^{coll}_{1\mu}+M^{sp}_{1\mu}.\nonumber\\
\end{eqnarray}
where
\begin{equation}
Q_0=\frac{3}{4\pi}ZR_0^2\beta\cos\gamma,\;\;
Q_2=\frac{3}{4\pi}ZR_0^2\beta\sin\gamma/\sqrt{2}.
\end{equation}
The relative sign of the two terms involved in the expression of the E2 transition operator is compatible with the structure of the moments of inertia ${\cal J}^{rig}$ as well as with that of the single particle potential given by Eq. (3.1). Z and $R_0$ denote the nuclear charge and radius respectively, while $e_{eff}$ is the effective charge, which for $^{163}$Lu, is taken equal to 1.3. Standard notations are used for  the nuclear magneton ($\mu_N$),the gyromagnetic factors of the rigid rotor ($G_R=Z/A$) and single particle characterizing the  orbital angular momentum 
($g_l$) and the spin ($g_s$), respectively. The angular momenta involved in Eq(3.38) are defined in the intrinsic frame of reference and transformed to the laboratory frame by means of the rotation matrices $D^{2}_{\mu\nu}$ and  $D^1_{\mu\nu}$, respectively.
The wave function for the states from the TSD1 band originates from the trial function employed in the time dependent variational principle:
\begin{eqnarray}
\Psi_{IM;j}&=&Ne^{zI_-}e^{sj_-}|IMI\rangle|jj\rangle \nonumber\\
           &=&\sum_{K,\Omega}\frac{z^{I-K}s^{j-\Omega}}{(1+|z|^2)^I(1+|s|^2)^j}
\left(\begin{matrix}2I \cr I-K\end{matrix}\right)^{1/2}
\left(\begin{matrix}2j \cr j-\Omega\end{matrix}\right)^{1/2}|IMK\rangle|j\Omega\rangle.
\label{funpsi}
\end{eqnarray}
Here $|IMK\rangle$ is the normalized Wigner function, $\sqrt{\frac{2I+1}{8\pi^2}}D^{I}_{MK}$.
Expressing the variables $z$ and $s$ in terms of the creation ($a^{\dagger};b^{\dagger}$) and annihilation operators ($a; b$) the wave function becomes and element of the Fock space spanned by
the states
\begin{equation}
|m,n\rangle_I =\frac{a^{\dagger\;m}b^{\dagger\;n}}{\sqrt{m!n!}}|0\rangle_I,
\end{equation}
with $|0\rangle_I$ standing for the vacuum state for the boson operators $a$ and $b$.
In our formalism, the states of the TSD1 band are described by the function (\ref{funpsi}) considered in the minimum point of the classical energy function:
\begin{equation}
\Psi_{IM;j}\left|_{\begin{matrix}(\varphi,r)=(0,I) \cr (\psi,t)=(0,j)\end{matrix}}\right.=
\frac{1}{2^{I+j}}\sum_{K,\Omega}\left(\begin{matrix}2I \cr I-K\end{matrix}\right)^{1/2}
\left(\begin{matrix}2j \cr j-\Omega\end{matrix}\right)^{1/2}|IMK\rangle|j\Omega\rangle|0\rangle_I.
\end{equation}
\begin{table}
\begin{tabular}{|c|c|cc|cc|}
\hline
&&\multicolumn{2}{c|}{$B(E2;I^+n_w\to (I-2)^+n^{'}_w)$} &\multicolumn{2}{c|}{$Q_I$}\\

&&\multicolumn{2}{c|}{$[e^2b^2]$}    &\multicolumn{2}{c|}{$[b]$}  \\
\hline
&&\multicolumn{2}{c|}{$n_w=0\;n^{'}_w=0$} &\multicolumn{2}{c|}{ $n_w=0\;n^{'}_w=0$}\\
TSD1&$I^{\pi}$&Th.  &  Exp.& Th. &Exp. \\
\hline
&$\frac{41}{2}^+$&2.72&3.45$^{+0.80}_{-0.69}$&9.33&9.93$^{+1.14}_{-0.99}$\\
&$\frac{45}{2}^+$&2.75&3.07$^{+0.48}_{-0.43}$&9.36&9.34$^{+0.72}_{-0.65}$\\
&$\frac{49}{2}^+$&2.77&2.45$^{+0.28}_{-0.25}$&9.40&8.32$^{+0.47}_{-0.42}$\\
&$\frac{53}{2}^+$&2.79&2.84$^{+0.24}_{-0.22}$&9.42&8.93$^{+0.38}_{-0.35}$\\
&$\frac{57}{2}^+$&2.80&2.50$^{+0.32}_{-0.29}$&9.43&8.37$^{+0.54}_{-0.49}$\\
&$\frac{61}{2}^+$&2.82&1.99$^{+0.26}_{-0.23}$&9.47&7.45$^{+0.49}_{-0.43}$\\
&$\frac{65}{2}^+$&2.83&1.95$^{+0.44}_{-0.30}$&9.48&7.37$^{+0.82}_{-0.57}$\\
&$\frac{69}{2}^+$&2.84&2.10$^{+0.80}_{-0.48}$&9.50&7.63$^{+1.46}_{-0.88}$\\
\hline
TSD2&            &\multicolumn{2}{c|}{$n_w=1\;n^{'}_w=1$} &\multicolumn{2}{c|}{ $n_w=1\;n^{'}_w=1$}\\
    &            &   Th.       &   Exp.   &  Th.   &  Exp.\\
\hline
&$\frac{47}{2}^+$&2.76&2.56$^{+0.57}_{-0.44}$&9.38&8.51$^{+0.95}_{-0.73}$\\
&$\frac{51}{2}^+$&2.78&2.67$^{+0.41}_{-0.33}$&9.41&8.67$^{+0.66}_{-0.53}$\\
&$\frac{55}{2}^+$&2.80&2.81$^{+0.53}_{-0.41}$&9.43&8.88$^{+0.83}_{-0.64}$\\
&$\frac{59}{2}^+$&2.81&2.19$^{+0.94}_{-0.65}$&9.46&7.82$^{+1.66}_{-1.15}$\\
&$\frac{63}{2}^+$&2.82&2.25$^{+0.75}_{-0.48}$&9.47&7.91$^{+1.32}_{-0.84}$\\
&$\frac{67}{2}^+$&2.83&1.60$^{+0.52}_{-0.37}$&9.49&6.66$^{+1.09}_{-0.76}$\\
&$\frac{71}{2}^+$&2.84&1.61$^{+0.82}_{-0.49}$&9.51&6.68$^{+1.70}_{-1.02}$\\
\hline
\end{tabular}
\caption{The E2 intra-band transitions $I\to (I-2)$ for TSD1 and TSD2 bands are listed. Also, the transition quadrupole moments are given. Theoretical results (Th.) are compared with the corresponding experimental data (Exp.) taken from Ref. \cite{Gorg}. B(E2) values are given in units of $e^2b^2$,
while the quadrupole transition moment, in $b$.}
\end{table}

\begin{table}
\begin{tabular}{|c|cc|cc|cc|}
\hline
&\multicolumn{2}{c|}{$B(E2;I^+n_w\to (I-1)^+n^{'}_w)$} &\multicolumn{2}{c|}{$B(M1;I^+n_w\to (I-1)^+n^{'}_w)$}&\multicolumn{2}{c|}{$\delta_{I\to(I-1)}$}\\

&\multicolumn{2}{c|}{$[e^2b^2]$}&\multicolumn{2}{c|}{$[\mu_N^2]$}&\multicolumn{2}{c|}{$[MeV.fm]$}  \\
&\multicolumn{2}{c|}{$n_w=1\;n^{'}_w=0$} &\multicolumn{2}{c|}{ $n_w=1\;n^{'}_w=0$}&\multicolumn{2}{c|}{$n_w=1\;n^{'}_w=0$}\\
$I^{\pi}$&Th.  &  Exp.& Th. &Exp. &Th.&Exp.\\
\hline
$\frac{47}{2}^+$&0.60&0.54$^{+0.13}_{-0.11}$&0.011&0.017$^{+0.006}_{-0.005}$&-2.7&-3.1$^{+0.36}_{-0.44}$\\
$\frac{51}{2}^+$&0.65&0.54$^{+0.09}_{-0.08}$&0.013&0.017$^{+0.005}_{-0.005}$&-2.7&-3.1$\pm 0.4$$^{a)}$\\
$\frac{55}{2}^+$&0.70&0.70$^{+0.18}_{-0.15}$&0.015&0.024$^{+0.008}_{-0.007}$&-2.8&-3.1$\pm 0.4$$^{a)}$\\
$\frac{59}{2}^+$&0.74&0.65$^{+0.34}_{-0.26}$&0.017&0.023$^{+0.013}_{-0.011}$&-2.8&-3.1$\pm 0.4$$^{a)}$\\
$\frac{63}{2}^+$&0.79&0.66$^{+0.29}_{-0.24}$&0.020&0.024$^{+0.012}_{-0.010}$&-2.8&\\
\hline
\end{tabular}
\caption{The B(E2) and B(M1) values for the transitions from TSD2 to TSD1.Mixing ratios are also mentioned. Theoretical results (Th.) are compared with the corresponding experimental (Exp.) data taken from Ref.\cite{Gorg}. Data labeled by $^{a)}$ are from Ref.\cite{Reich}.}
\end{table}
Concerning the excited bands TSD2, TSD3, TSD4, they are described by the wobbling excited states:
\begin{equation}
\Phi_{I+n_w,M+n_w;jn_w}=\Psi_{IM;j}\left|_{\begin{matrix}(\varphi,r)=(0,I) \cr (\psi,t)=(0,j)\end{matrix}}
\frac{1}{\sqrt{n_w!}}\left(\Gamma^+_1\right)^{n_w}|0)_I\right.
\end{equation}
for $n_w=1,2,3$ respectively. The phonon operator $\Gamma^+_1$ corresponds to the wobbling energy $\Omega_1$. The vacuum state for the wobbling phonons is denoted by $|0)_I$. Its expression as well as its overlap with the vacuum state $|0\rangle_I$ for the operators $a$ and $b$ are given in Appendix D. The angular momentum and its projection on the OZ axis accompanying the wave function9n $\Phi$ are obtained by considering the rotation properties of the phonon operator $\Gamma^+_1$, specified above.

The reduced E2 and M1 transition probabilities have the expressions:
\begin{eqnarray}
B(E2;I_i^{\pi};jn_w\to I_f^{\pi};jn^{\prime}_w)&=&\langle I_i^{\pi};jn_w||{\cal M}(E2)||I_f^{\pi};jn^{\prime}_w\rangle ^2,\nonumber\\
B(M1;I_i^{\pi};jn_w\to I_f^{\pi};jn^{\prime}_w)&=&\langle I_i^{\pi};jn_w||{\cal M}(M1)||I_f^{\pi};jn^{\prime}_w\rangle ^2.
\end{eqnarray}
Note that the reduced matrix elements are defined according to the Rose convention \cite{Rose}:
\begin{equation}
\langle JM|T_{k\mu}|J'M'\rangle=C^{J'\;k\;J}_{M'\;\mu\;M}\langle J||T_{k}||J'\rangle.
\end{equation}
The reduced matrix elements for the electric and magnetic transition operators have the analytical expressions given in Appendix D. They are also used for calculating the mixing ratio for the E2 and M1 transitions from the TSD2 to TSD1 bands, defined as \cite{Toki1,Kran}:
\begin{equation}
\delta=8.78\times 10^{-4}E_{if}\frac{\langle I_i||{\cal M}(E2)||I_f\rangle}{\langle I_i||{\cal M}(M1)||I_f\rangle}.
\end{equation}
The matrix element for the E2 transition is taken in units of $e\cdot fm^2$, while that for the M1 transition in $e\cdot fm$. The transition energy is denoted by $E_{if}$ and is taken in $MeV$.
We also calculated the transition quadrupole moments, according to the definition \cite{Hage84}:
\begin{equation}
Q_I=\sqrt{\frac{16\pi}{5}}\langle I||{\cal M}(E2)||I-2\rangle/
C^{I\;2\;I-2}_{K\;0K} .
\end{equation}
It turns out that $Q_I$ varies only slightly with K. In our calculations we considered K=1/2.
Results are collected in Tables II and III, where for comparison the available experimental data are also listed.
We note that the present results agree quite well with the data. Table II indicates a large value for the transition quadrupole moment which in fact is consistent with the large nuclear deformation of $^{163}$Lu. A large transition quadrupole moment in the band TSD1 of $^{163}$Lu was also pointed out in Refs.\cite{Gorg,Scho1,Schmi}. The calculated matrix elements involve an effective charge $e_{eff}=1.3$. In addition, for the inter-band transition matrix elements a common quenching factor $F_q$ was necessary. This accounts for the contribution of the multi boson ($a$ and $b$) states to the $\Delta I=1$ transitions. Indeed, by including the many boson components into the wave function, the norm decreases. The need of introducing a quenching factor is consistent with the fact that that collective states with low energy show strong inter-band transitions.The factor $F_q$ was chosen so that an overall agreement for the inter-band transitions is obtained. 
Thus, the adopted value is $F_q=0.1144$ for all transitions listed in Table III. One remarks on the large magnitude of the B(E2) and B(M1) values for the observed transitions from TSD2 to TSD1. This  is, indeed, one of the features specific to the wobbling motion. Note that the quenching factor does not affect the mixing ratios. Another specific feature of the present formalism is the small value for the wobbling frequency $\Omega_1$ which is used as elementary excitation for all four super-deformed bands. The soft character for the $\Omega_1$ excitation reclaims in fact a $\gamma$ softness since indeed in the limit of axial symmetry the mode energy is vanishing. This is however  consistent with the rigid structure of the moment of inertia.  From Figs. 19 and 20 we see that the softness prevails for low spin whereas by increasing I the rigidity is slightly raising its importance. Of course we can ask the question
whether the $\gamma$ rigidity hypothesis is realistic or not? Obviously, a dynamic character
of $\gamma$ together with the $\gamma$ dependence of the wave functions involved in the inter-band transition, would decrease the corresponding matrix elements. Such a feature might be also used
for justifying the presence of a quenching factor. 

It is worth noting that the excited bands TSD2, TSD3 and TSD4 are formed of wobbling multi-phonon
states built with the operator $\Gamma^{\dagger}_1$. We recall, however, the fact that there are two phonons which correspond to the precession of the two angular momenta ${\bf I}$ and $\bf j$, respectively. This suggests that, in principle, some other TSD bands may be defined by activating the phonon operator $\Gamma^{\dagger}_2$ with the wobbling frequency $\Omega_2$. Relevant data on this line would encourage us to pursue this project. 

As mentioned before the band TSD4 is considered to  have an wobbling nature and that agrees with the calculations from Ref.\cite{Tana3}. This picture requires, however, a further improvement in order to have an agreement with the predictions of Ref. \cite{Jens1}. Indeed, in the quoted reference one claims that the band TSD4 is of different nature as compared with TSD2,3 bands. This conclusion is based on the results for the dynamic moment of inertia as well as on the spin alignment. Indeed,  for the first two excited bands the dynamic moment of inertia exhibits a bump which reflects the spin alignment of the core particles, while in the case of TSD4 this bump is missing.
On the other hand the alignment for TSD4 is by $3\hbar$ larger than in the other cases. Based on microscopic calculations with ultimate cranking, it was concluded that TSD4 is determined by a three quasiparticle configuration. The observed large E1 transition connecting TSD4 and TSD1 could be however explained only by allowing an admixture of the mentioned configuration with an octupole vibration. The TSD potential exhibits two wells, one for the normal deformed bands and one for triaxial super-deformed bands. The barrier separating the two wells is quite consistent such that there is no transition between the TSD and the normal deformed bands. The band TSD4 belongs to the TSD well and therefore
the coexistence of a three quasiparticle band and three TSD bands of wobbling nature suggests that
in fact TSD4 has properties of both three quasiparticle and wobbling nature. 
Details about the extension of the present formalism as to be able to describe the E1 transitions to the yrast TSD1 will be presented in a forthcoming publication. Here wee just mention that including an octupole term in the dipole transition operator the transition from the TSD4 to the yrast TSD1 is possible.

Indeed, let us consider the dipole operator:
\begin{equation}
T_{1M}=\sum_{M_1,M_2,\mu}\left[\beta\cos\gamma D^{2}_{M_1,0}+\beta\frac{\sin\gamma}{\sqrt{2}}
(D^{2}_{M_1,2}+D^{2}_{M_1,-2})\right]\otimes D^{3}_{M_2,\mu}r^3Y_{3\mu} C^{2\;3\;1}_{M_1\;M_2\;M}.
\end{equation}
Obviously, the matrix elements of this operator between a $I$ state from the TSD4 band and a $I\pm 1$ state from the TSD1 band is nonvanishing. Moreover, the overlap of the three phonon component of the initial state and the Hartree-Fock  boson states involved in the chosen TSD1 band is also nonvanishing. This is proved in Appendix E, where the mentioned overlap is analytically given.

\section{Conclusions}
\renewcommand{\theequation}{4.\arabic{equation}}
\setcounter{equation}{0}
In the previous sections we presented a semi-classical formalism to describe the wobbling motion in
even-even and even-odd nuclei. In both cases one uses a time dependent variational principle in connection with a rotor and a particle-rotor system, respectively. In this way the quantum mechanical eigenvalue problem is transformed into a system of classical equations in a phase space.
Harmonic solutions are further quantized which results in getting a compact formula for the wobbling motion. Results corresponding to the three pictures, quantal, classical and re-quantal, are in detail compared for the even-even case. An excellent agreement with each other shows up for the yrast and the next two excited bands. The requantized equations leads to the Bargmann representation of the triaxial rotor which provides a compact expression for the wobbling frequency. Application to $^{158}$Er, indicates a quite good agreement with the available data for both energies and B(E2) values. 

Since in the even-odd case one deals with two angular momenta, the total (${\bf I}$) and the individual ({\bf j}) ones, one ends up with two wobbling frequencies. The lowest one is used to calculate the energies   in the bands $TSD1$, $TSD2$, $TSD3$ and $TSD4$, the intra-band and inter-band $E2$ and $M1$ reduced transition probabilities for $TSD2$ and $TSD1$.
The agreement with available data is also impressively good.

The variational state is a state of good angular momentum, but a mixture of components with different $K$ quantum number. This assures that states of different angular momenta are orthogonal and moreover provides a prerequisite of the wobbling motion where the projection of angular momentum on the axis OZ fluctuates around a static value. Also, despite the fact the trial function suggests that the quantization axis is OZ, the classical angular momentum is oriented along the OX axis. This is reflected in the structure of the trial function for the core system, its amplitudes being picked on the $K=0$ component, which agrees with the behavior of the exact eigenfunction as indicated by Fig.1. This feature makes the present formalism to be at par with the cranking models where this orientation for the angular momentum is obtained by the cranking constraint. However, for an even-odd system, the mechanism of coupling the core and odd particle angular momenta in the two pictures are different\cite{Hage}. Indeed, within the cranking formalism the core angular momentum is always oriented along the principal axis with the largest moment of inertia and the odd particle may rotate around an axis which do not coincide with the above mentioned  axis. On the other hand within the wobbling regime the core angular momentum can tilt apart from the axis with the largest moment of inertia, while the odd particle always rotates  around the mentioned axis. As discussed in Ref.\cite{Marsh} the wobbling of the core angular momentum is taken into account not in the mean field level, that is, the cranking model in a narrow sense, but in the RPA level in the cranking approach. Similar feature is met in the present formalism where in Eq. (3.37), ${\cal H}_{min}$  does not account for the core angular momentum wobbling, while at the RPA level the effect of wobbling motion shows up.

In the even-odd case there are two distinct phonon frequencies, both of them being involved in the zero point motion of the odd system which determines the yrast energies, i.e. the TSD1 band. However, the excited bands are determined  by the contribution of the $\Omega_1$ phonon and the zero point energy due to both wobbling frequencies.
We may ask ourself whether the $\Omega_2$ multiphonon states or a combination of the two phonons, $\Omega_1$ and $\Omega_2$, generate additional TSD bands.

It is worth mentioning that in both even-even and even-odd cases, the wobbling frequency is vanishing when two moments of inertia are equal.  Therefore, one could assert that the wobbling motion is a signature of the triaxiality. However, the low value of the wobbling frequency indicates a $\gamma$-soft regime. The behavior of $\Omega_1$ as function of angular momentum reflects the softness vs rigidity competition.

The wobbling motion in $^{163}$Lu has been also treated by several authors 
\cite{Ham,Ham1,Matsu,Tana,Tana3,Oi,Alme,Tana4,Frau} by using different methods. Our results are consistent with those of the quoted references.

Finally, we conclude that the present formalism seems to be an efficient tool to describe quantitatively the wobbling motion in even-even and even-odd nuclei. Application of the formalism to other nuclei, from different area of nuclear chart, will be the objective of a forthcoming paper.

\section{Appendix A}
\renewcommand{\theequation}{A.\arabic{equation}}
\setcounter{equation}{0}
Here we give the analytical expressions for the matrix elements $A_{ij}$ and $B_{ij}$.
\begin{eqnarray}
A_{11}&=&\frac{1}{I}\left[(j+1-2I)A_1+(2I-1)A_3\right],\nonumber\\
A_{12}&=&A_{21}=-A_3,\nonumber\\
A_{22}&=&\frac{1}{j}\left((I+1-2j)A_1+(2j-1)A_3+V\frac{2j-1}{j(j+1)}\sqrt{3}(\sqrt{3}\cos\gamma+
\sin\gamma)\right),\nonumber\\
B_{11}&=&-I\left[(2I-1)(A_2-A_1)+jA_1\right],\nonumber\\
B_{12}&=&B_{21}=IjA_2,\nonumber\\
B_{22}&=&-\left(j(2j-1)(A_2-A_1)+V\frac{2j-1}{j+1}2\sqrt{3}\sin\gamma+IjA_1\right).
\end{eqnarray}

\section{Appendix B}
\renewcommand{\theequation}{B.\arabic{equation}}
\setcounter{equation}{0}
The wobbling phonon amplitudes can be analytically obtained by solving the random phase approximation (RPA)-like equations.  The result is as follows:
\begin{eqnarray}
X_1&=&\frac{\left(A_3kk'-Ij\frac{A_1}{kk'}\right)\omega_2}{A_2A_3Ij-(\omega_1-\Omega)(\omega_2-\Omega)}Y_2,\nonumber\\
X_2&=&\left[\frac{2\omega_2(\omega_1-\Omega)}{A_aA_3Ij-(\omega_1-\Omega)(\omega_2-\Omega)}+1\right]\frac{A_3kk'-Ij\frac{A_2}{kk'}}{A_3kk'+Ij\frac{A_2}{kk'}}Y_2,\nonumber\\
Y_1&=&\left[\omega_2+\Omega+\frac{\frac{1}{2}\omega_2\left(A_3kk'-Ij\frac{A_2}{kk'}\right)^2}
{A_2A_3Ij-(\omega_1-\Omega)(\omega_2-\Omega)}\right]\frac{2Y_2}{A_3kk'+Ij\frac{A_2}{kk'}},
\nonumber\\ 
{Y_2}^{-1}&=&\frac{2}{A_3kk'+Ij\frac{A_2}{kk'}}\left\{-A_2A_3Ij-(\omega_2+\Omega)^2+
\frac{\omega_2(A_3kk'-Ij\frac{A_2}{kk'})^2}{A_2A_3Ij-(\omega_1-\Omega)(\omega_2-\Omega)}\right.
\nonumber\\
&\times&\left.\left[\frac{\omega_2\left(A_2A_3Ij+(\omega_1-\Omega)^2\right)}{A_2A_3Ij-(\omega_1-\Omega)(\omega_2-\Omega)}
+(\omega_1-\omega_2-2\Omega)\right]\right\}^{1/2}.
\end{eqnarray}
\section{Appendix C}
\renewcommand{\theequation}{C.\arabic{equation}}
\setcounter{equation}{0}
The reduced matrix elements for the electric quadrupole transition operator have the following analytical expressions:
\begin{eqnarray}
&&\langle \Phi_{I;jn_w}||T^{coll}_{2}||\Phi_{I';jn^{\prime}_{w}}\rangle =e_{eff}\frac{3}{4\pi}ZR_0^2\frac{\hat{I}'}{\hat{I}}\sum_{\begin{matrix}K,\;K'\;~\cr\Omega,\;\Omega';\nu\end{matrix}}
{\cal N}_{I'}F_{n_wn^{\prime}_w}(I',K',\Omega')
C_{IK}C_{j\Omega}C_{j\Omega'}\nonumber\\&&\times\left[\beta\cos{\gamma}C_{I'K}
C^{I'\;2\;I}_{K\;0\;K}-\frac{1}{\sqrt{2}}\beta\sin{\gamma}\left(C_{I'K-2}C^{I'\;2\;I}_{K-2\;2\;K}+C_{I'K+2}C^{I'\;2\;I}_{K+2\;-2\;K}\right)\right],\nonumber\\
&&\langle \Phi_{I;jn_w}||T^{sp}_{2}||\Phi_{I;jn^{\prime}_{w}}\rangle =e_{eff}
\sqrt{\frac{5}{4\pi}}\frac{\hbar}{M\omega}(N+\frac{3}{2})C^{j\;2\;j}_{\frac{1}{2}\;0\;\frac{1}{2}}
\frac{\hat{I}'}{\hat{I}}\nonumber\\
&&\sum_{\begin{matrix}K,\;K'\;~\cr\Omega,\;\Omega';\nu\end{matrix}}
C_{IK}C_{I'K'}C_{j\Omega}C_{j\Omega'}
C^{I'\;2\;I}_{K'\;\nu\;K}C^{j\;2\;j}_{\Omega'\;\nu\;\Omega}F_{n_wn^{\prime}_w}(I',K',\Omega')
{\cal N}_{I'}.
\end{eqnarray}
where ${\cal N}_{I'}$ stands for the first order expanded state $I'$.
The wave function amplitudes were denoted by:
\begin{equation}
C_{IK} = \frac{1}{2^I}\left(\begin{matrix}2I\cr I-K\end{matrix}\right)^{1/2},\;\;
C_{j\Omega} =\frac{1}{2^j}\left(\begin{matrix}2j\cr j-\Omega\end{matrix}\right)^{1/2}.
\end{equation}
The matrix elements corresponding to $n_w=1$ and $n'_w=0$ were calculated by expanding in the first order the  function$\Psi$ around the minimum point of the classical energy function. Indeed this is the leading term in the overlap matrix element of  one phonon state and a state from the TSD1 band, and has the form:
\begin{eqnarray}
F_{10}(K',\Omega')&=&\frac{i}{\sqrt{2}}\left[\frac{K'}{I'}k_{I'}(X_1-Y_1)+\frac{I'-K'}{k_{I'}}
(X_1+Y_1)\right.\nonumber\\
&+&\left.\frac{\Omega'}{j}k'_{I'}(X_2-Y_2)+\frac{j-\Omega'}{k'_{I'}}(X_2+Y_2)\right]{_I}
(0|0\rangle_{I'}F_q.
\end{eqnarray}
The other overlaps can be easily calculated with the result:
\begin{eqnarray}
F_{00}&=&F_{11}=F_{22}=1,\nonumber\\
F_{21}&=&\sqrt{2}F_{10}.
\end{eqnarray}
The quenching factor $F_q$ was taken equal to 0.1144 for all transitions of both E2 and M1 type.
The overlap $_I(0|0 \rangle_{I'}$ is explicitly given in appendix D. 
As for the magnetic transition operator the reduced  matrix elements are:
\begin{eqnarray}
\langle I;jn_w||M^{coll}_1||I';jn'_w\rangle &=& \mu_Ng_R\sqrt{\frac{3}{4\pi}}\sqrt{I'(I'+1)}\frac{\hat{I'}}{\hat {I}}
\sum_{K,K'}C_{IK}C_{I'K'}C_{j\Omega}C_{j\Omega'}{\cal N}_{I'}\nonumber\\
&\times&F_{n_wn'_w}(I',K',\Omega')C^{I'\;1\;I}_{K'-\nu\;\nu K}C^{I'\;1\;I'}_{K'-\nu\;\nu\; K'-\nu},\nonumber\\
\langle I;jn_w||M^{sp}_1||I';jn'_w\rangle &=&\mu_N\sqrt{\frac{3}{4\pi}}\frac{\hat{I'}}{\hat {I}}
\sum_{\begin{matrix}K,\;K'\;~\cr\Omega,\;\Omega';\nu\end{matrix}}C_{IK}C_{I'K'}C_{j\Omega}
C_{j\Omega'}F_{n_wn'_w}(I',K',\Omega'){\cal N}_{I'}\nonumber\\
\times C^{I'\;1\;I}_{K'\;\nu\;K}C^{j\;1\;j}_{\Omega'\;\nu\;\Omega}
&&\left[(g_l-g_R)[j(j+1)]^{1/2}+\frac{g_s-g_l}{2}\frac{\frac{3}{4}+j(j+1)-l(l+1)}{[j(j+1)]^{1/2}}\right]
\end{eqnarray}
\section{Appendix D}
\renewcommand{\theequation}{D.\arabic{equation}}
\setcounter{equation}{0}
For what follows it is convenient to introduce the new bosons:
\begin{eqnarray}
\gamma^{\dagger}_a &=&\bar{X}_1a^{\dagger}-\bar{Y}_1a,\;
\gamma^{\dagger}_b =\bar{X}_2b^{\dagger}-\bar{Y}_2b,\;\;\rm{with}\nonumber\\
\bar{X}_1&=&\frac{X_1}{\sqrt{X_1^2-Y_1^2}},\;\bar{Y}_1=\frac{Y_1}{\sqrt{X_1^2-Y_1^2}},\nonumber\\
\bar{X}_2&=&\frac{X_2}{\sqrt{X_2^2-Y_2^2}},\;\bar{Y}_2=\frac{Y_2}{\sqrt{X_2^2-Y_2^2}},
\end{eqnarray}

The vacuum state for the phonon operator $\Gamma^{\dagger}_1$, denoted by $|0)_I$ is related to the vacuum state of the boson operators $a$ and $b$, $|0\rangle_I$ by the transformation:
\begin{equation}
|0)_I=e^{x(a^{\dagger\;2}-a^2)+y(b^{\dagger\;2}-b^2)}|0\rangle _I,
\end{equation}
where the parameters $x$ and $y$ are related with the amplitudes $(\bar{X}_1;\bar{X}_1)$ and 
$(\bar{X}_2;\bar{Y}_2)$ through the equation:
\begin{eqnarray}
\cosh(2x)&=&\bar{X}_1,\;\;\sinh(2x)=\bar{Y}_1,\nonumber\\
\cosh(2y)&=&\bar{X}_2,\;\;\sinh(2y)=\bar{Y}_2.
\end{eqnarray}
Indeed, one can easily prove that:
\begin{equation}
\Gamma_1|0)_I=0.
\end{equation}
The excited states of the two vacuua are:
\begin{eqnarray}
|m)_I &=&\frac{1}{\sqrt{m!}}\left(\Gamma^{\dagger}\right)^m|0)_I
=\frac{1}{\sqrt{m!}}\sum_{i}\left(\begin{matrix}m\cr i\end{matrix}\right)
\left(X_1a^{\dagger}-Y_1a\right)^i\left(X_2b^{\dagger}-Y_2b\right)^{m-i}|0)_I
\;m=1,2,3,...\nonumber\\
|n,k\rangle_I &=&\frac{1}{\sqrt{n!k!}}a^{\dagger\;n}b^{\dagger\;k}|0\rangle_I,\;n,k=1,2,3,....
\end{eqnarray} 
Thus, the multi-phonon state becomes:
\begin{equation}
|m)_I=\frac{1}{\sqrt{m!}}\sum_{i}\left(\begin{matrix}m\cr i\end{matrix}\right)
\left(X_1^2-Y_1^2\right)^{i/2}\left(X_2^2-Y_2^2\right)^{(m-i)/2}(\gamma^{\dagger}_a)^i
(\gamma^{\dagger}_b)^{m-i}|0)_I.
\end{equation}
Its overlap matrix elements with the state $|n,k\rangle_I$ is:
\begin{equation}
_{I}(m|n,k\rangle_{I}=\frac{1}{\sqrt{m!}}\sum_{i}\left(\begin{matrix}m\cr i\end{matrix}\right)
\left(X_1^2-Y_1^2\right)^{i/2}\left(X_2^2-Y_2^2\right)^{(m-i)/2}\sqrt{i!(m-i)!}G^{a}_{in}(x)
G^{b}_{m-i,k}(y),
\end{equation}
where the matrices $G^{s}_{mk}$ with $s=a,b$ are the overlaps:
\begin{eqnarray}
G^{a}_{in}&=&\frac{1}{\sqrt{i!n!}} {_{I}}(0|\gamma_a^ia^{\dagger\;n}|0\rangle_I,\nonumber\\
G^{b}_{m-i,k}&=&\frac{1}{\sqrt{(m-i)!k!}} {_{I}}(0|\gamma_b^{\dagger\;(m-i)}b^{\dagger\;k}|0\rangle_I.
\end{eqnarray}
This type of matrix was analytically calculated by one of us  (A. A. R.) in Ref.\cite{Rad77}, with the result:
\begin{eqnarray}
G^{s}_{n,m}(y)&=&\sqrt{m!n!}(\cosh y)^{-(m+n+1)/2}\nonumber\\
&\times&\sum_{q}\frac{(-1)^{(n-m)/2}}{q![(n-q)/2]![(m-q)/2]!}(\frac{1}{2}\sinh y)^{(m+n)/2-q},\nonumber\\
\rm{for}\;\;s&=&a,\;\;\cosh y =\bar{X}_1;\;\;\sinh y =\bar{Y}_1,\;\;\rm{while}\nonumber\\
\rm{for}\;\;s&=&b,\;\;\cosh y =\bar{X}_2;\;\;\sinh y =\bar{Y}_2.
\end{eqnarray}
In particular, for the overlap of interest one gets:
\begin{equation}
_{I}(0|0\rangle_{I}\equiv_{I}(0|0,0\rangle_{I}=\frac{\left[(X_1^2-Y_1^2)(X_2^2-Y_2^2)\right]^{1/4}}
{\left[X_1X_2\right]^{1/2}}.
\end{equation}
\section{Appendix E}
\renewcommand{\theequation}{E.\arabic{equation}}
\setcounter{equation}{0}
Here we shall present the quantized expression for the coefficients involved in the expansion (3.40):

\begin{equation}
f(z)=\frac{z^{I-K}}{(1+|z|^2)^I};\;\;g(s)=\frac{s^{j-\Omega}}{(1+|s|^2)^j}. 
\end{equation}
We explain the procedure for the case of $f(z)$ and express this function in terms of the canonical variables $(r,\phi)$ which are related with the corresponding energy minimum point by:
\begin{equation}
r=I+r^{'};\;\; \varphi=\varphi^{'}.
\end{equation}
With the new variables the function acquires the form:
\begin{equation}
f(r',\phi')=\frac{1}{(2I)^I}(I+r^{'})^{\frac{I+K}{2}}(I-r^{'})^{\frac{I-K}{2}}e^{i(I-K)\varphi^{'}}.
\end{equation}
The factor depending on the momentum $r'$ is expanded  up to the second order, while the one depending on the coordinate $\phi'$ is treated without any approximation.
The resulting expression is first symmetrized and then quantized by the replacement:
\begin{equation}
r'=\frac{ik}{\sqrt{2}}(a^{\dagger}-a);\;\;\varphi'=\frac{1}{\sqrt{2}k}(a^{\dagger}+a).
\end{equation}
One proceeds in a similar way with the function $g(s)$. The final result for the quantized form
of the product $fg$ is:
\begin{eqnarray}
&&f(z)g(s)=\frac{1}{2^{I+j}}e^{-\frac{1}{4}\left(\frac{I-K}{k}\right)^2}
e^{-\frac{1}{4}\left(\frac{j-\Omega}{k'}\right)^2}\sum_{m,n}i^{m+n}\left(\frac{I-K}{\sqrt{2}k}\right)^n\left(\frac{j-\Omega}{\sqrt{2}k'}\right)^m\frac{1}{\sqrt{n!m!}}
\nonumber\\
&\times&A_0B_0|nm\rangle_I+A_0B_1\sqrt{m+1}|n,m+1\rangle_I +A_0B_2\sqrt{(m+1)(m+2)}|n,m+2\rangle_I 
\nonumber\\
&+&A_1B_0\sqrt{n+1}|n+1,m\rangle_I +A_1B_1\sqrt{(n+1)(m+1)}|n+1,m+1\rangle_I 
\nonumber\\&+&A_1B_2\sqrt{(n+1)(m+1)(m+2)}|n,m+2\rangle_I 
\nonumber\\
&+&A_2B_0\sqrt{(n+1)(n+2)}|n+2,m\rangle_I +A_2B_1\sqrt{(n+1)(n+2)(m+1)}|n+2,m+1\rangle_I 
\nonumber\\&+&A_2B_2\sqrt{(n+1)(n+2)(m+1)(m+2)}|n+2,m+2\rangle_I,
\end{eqnarray}
where the states $|n,m\rangle$ are those defined by Eq. (3.41), while  the factors $A_m, \;B_m$ with m=0,1,2, have the following expressions:
\begin{eqnarray}
A_0&=&1-\frac{K(I-K)}{I}+\frac{13}{24I^2}(K^2-I)(I-K)^2+\frac{k^2}{4I^2}(K^2-I),\nonumber\\
A_1&=&\frac{ik}{\sqrt{2}}\left(\frac{K}{I}-\frac{1}{I^2}(K^2-I)(I-K)\right),\nonumber\\
A_2&=&-\frac{k^2}{4I^2}(K^2-I),\nonumber\\
B_0&=&1-\frac{\Omega(j-\Omega)}{j}+\frac{13}{24j^2}(\Omega^2-j)(j-\Omega)^2+\frac{k^{'2}}{4j^2}(\Omega^2-j),\nonumber\\
B_1&=&\frac{ik'}{\sqrt{2}}\left(\frac{\Omega}{j}-\frac{1}{j^2}(\Omega^2-j)(j-\Omega)\right),
\nonumber\\
B_2&=&-\frac{k^{'2}}{4j^2}(\Omega^2-j).
\end{eqnarray}
It is clear now that the overlap of the  three phonon state $|3)_I$(D.5) and the function $f(z)g(s)$ (E.5) is a superposition of the partial overlaps $_{I}(3|n,m\rangle_I$, which were analytically expressed in Appendix D, Eq. (D.7).
Thus, the statement concerning the existence of a non-vanishing transition matrix element connecting the bands TSD4 and TSD1 is completely proved.

\end{document}